\documentclass[runningheads]{llncs}

\usepackage{listings}
\usepackage{graphicx}
\usepackage{xcolor}
\usepackage{comment}
\usepackage[hidelinks]{hyperref}

\usepackage{tikz}
\usetikzlibrary{patterns}

\tikzstyle{textc}=[anchor=south]
\tikzstyle{textl}=[anchor=east, color=black!50!green]
\tikzstyle{textr}=[anchor=west, color=orange]

\tikzstyle{cpu}=[draw, rectangle, anchor=north, fill=green!50!black, minimum width=1.2cm]
\tikzstyle{other}=[draw, rectangle, anchor=north, fill=green!80!black, minimum width=1.2cm]
\tikzstyle{wait}=[draw, rectangle, anchor=north, fill=cyan!50!white, minimum width=1.2cm]
\tikzstyle{comp}=[draw, rectangle, anchor=north, pattern=north west lines, pattern color=magenta!80!black, minimum width=1.2cm]
\tikzstyle{nic}=[draw, dashed, rectangle, anchor=north, fill=orange!20!white, minimum width=0.8cm]

\newcommand{\subabc}[1]{
    \node[textc] at (0, #1) {\Large (a)};
    \node[textc] at (3, #1) {\Large (b)};
    \node[textc] at (6, #1) {\Large (c)};
}

\newcommand{\subab}[1]{
    \node[textc] at (0, #1) {\Large (a)};
    \node[textc] at (6, #1) {\Large (b)};
}

\newcommand{\legendA}[2]{
  \draw[dashed] (-1.5, #1 -0.8) rectangle (8.8, #1 +0.3);
  \node[draw, rectangle, anchor=north, fill=green!50!black, minimum width=0.5cm, minimum height=0.5cm] at (-0.5, #1){};
  \node[draw, rectangle, anchor=north, fill=cyan!50!white, minimum width=0.5cm, minimum height=0.5cm] at (2.5, #1){};
  \node[draw, dashed, rectangle, anchor=north, fill=orange!20!white, minimum width=0.5cm, minimum height=0.5cm] at (6, #1){};
  \node[anchor=west] at (-0.2, #1 -0.3) {\Large CPU};
  \node[anchor=west] at (2.8, #1 -0.3) {\Large #2};
  \node[anchor=west] at (6.3, #1 -0.3) {\Large NIC};
}

\newcommand{\legend}[2]{
  \draw[dashed] (-1.5, #1 -0.8) rectangle (8.8, #1 +0.3);
  \node[draw, rectangle, anchor=north, fill=green!50!black, minimum width=0.5cm, minimum height=0.5cm] at (-0.5, #1){};
  \node[draw, rectangle, anchor=north, fill=cyan!50!white, minimum width=0.5cm, minimum height=0.5cm] at (1.5, #1){};
  \node[draw, rectangle, anchor=north, pattern=north west lines, pattern color=magenta!80!black, minimum width=0.5cm, minimum height=0.5cm] at (3.7, #1){};
  \node[draw, dashed, rectangle, anchor=north, fill=orange!20!white, minimum width=0.5cm, minimum height=0.5cm] at (7, #1){};
  \node[anchor=west] at (-0.2, #1 -0.3) {\Large CPU};
  \node[anchor=west] at (1.8, #1 -0.3) {\Large #2};
  \node[anchor=west] at (4.0, #1 -0.3) {\Large Compute};
  \node[anchor=west] at (7.3, #1 -0.3) {\Large NIC};
}

\lstdefinestyle{proto}{
    basicstyle=\scriptsize,
    frame=single,
    language=C,
    framesep=5pt,
    breaklines=true,
    linewidth=0.95\textwidth,
    xleftmargin=.05\textwidth
}
\lstdefinestyle{example}{
    basicstyle=\scriptsize,
    language=C,
    frame=tb,
    showstringspaces=false,
    breaklines=true
}

\newcommand{\wait}{\textit{wait}\ }
\newcommand{\pollfn}{\texttt{poll\_fn}}

\begin{document}

\title{MPI Progress For All}

\author{Hui Zhou\inst{1} \and
Robert Latham\inst{1} \and
Ken Raffenetti\inst{1} \and
\\Yanfei Guo\inst{1} \and
Rajeev Thakur\inst{1}}

\institute{Argonne National Laboratory, Lemont, IL 60439, USA}
\maketitle
\begin{abstract}
The progression of communication in the Message Passing Interface (MPI) is not well defined, yet it is critical for application performance, particularly in achieving effective computation and communication overlap. The opaque nature of MPI progress poses significant challenges in advancing MPI within modern high-performance computing practices.
First, the lack of clarity hinders the development of explicit guidelines for enhancing computation and communication overlap in applications. Second, it prevents MPI from seamlessly integrating with contemporary programming paradigms, such as task-based runtimes and event-driven programming. Third, it limits the extension of MPI functionalities from user space.
In this paper, we examine the role of MPI progress by analyzing the implementation details of MPI messaging. We then generalize the asynchronous communication pattern and identify key factors influencing application performance. Based on this analysis, we propose a set of MPI extensions designed to enable users to construct and manage an efficient progress engine explicitly.
We compare our approach to previous efforts in the field, highlighting its reduced complexity and increased effectiveness.
\end{abstract}

\pagestyle{plain}
\section{Introduction}
% MPI progress is ambiguous about performance
Overlapping computation and communication \cite{castillo2019optimizing,sergent2018efficient} is a key performance goal in high-performance computing (HPC).
Ideally, with 100\% computation/communication overlap, communication and synchronization become effectively free, allowing parallel applications to scale perfectly.
However, achieving this overlap goal remains challenging.
The Message Passing Interface (MPI), the \textit{de facto} communication runtime for HPC applications, does not precisely define how communication progress is made.
MPI guarantees that once communication is initiated, it will complete, but it does not specify whether progress occurs during the starting call (\textit{e.g.,} \texttt{MPI\_Isend}), the completion call (\textit{e.g.,} \texttt{MPI\_Wait}), or in between.
To achieve effective computation/communication overlap, strong progress from MPI is desirable~\cite{holmes2020mpi}, meaning that 
  MPI can make progress between the starting and completion calls without explicit MPI calls from the user.
However, implementing strong progress poses constraints on MPI implementations and is not always feasible.
One approach is to employ a default asynchronous progress thread~\cite{ruhela2019efficient}. However, due to missing context from the application, a  global progress thread often leads to performance issues and is generally not considered a good solution.
Thus, it remains important for applications to have a strategy for managing MPI progress to achieve optimal performance. 

% MPI Request is not effective in managing progress
Currently, applications have limited means to explicitly control MPI progress.
Using an MPI blocking call ensures that the operation completes before the call returns, but it blocks the calling thread from performing computations.
With nonblocking MPI calls, an application can split an MPI operation into a start and a completion call and perform computation in between.
However, depending on the implementation, the communication may be blocked until the completion call rather than overlapping with the computation~\cite{hoefler2008message}.
To improve the communication/computation overlap, there needs to be a strategy to invoke MPI progress regularly during computation.
The current method for the application to invoke MPI progress is via \texttt{MPI\_Test}, but \texttt{MPI\_Test} is tied to a specific MPI request.
Thus, designing a progress engine that includes MPI progress requires a synchronization mechanism for managing MPI requests, which is often complex and prone to inefficiency. 

% The factors for high performance MPI Progress applies to asynchronous programming in general
% Modern event-driven programming
Asynchronous programming, including task-based  \cite{schuchart2021callback} and event-driven programming \cite{haller2006event}, is gaining popularity as modern computing architectures incorporate more cores and adopt more hybrid architectures.
Asynchronous task-based programming allows programmers to separate writing tasks from managing deployment and progression.
Event-driven programming alleviates the need to manage asynchronous handles and enables programmers to express tasks that involve multiple asynchronous events.
Compared to traditional MPI programming, where programmers need to partition jobs from the start and explicitly manage synchronization, these asynchronous programming styles significantly reduce the complexity of parallel programming.
However, MPI poses challenges to working with these asynchronous programming styles.
Task-based runtimes need to manage MPI nonblocking operations via MPI requests and regularly test for completion to maintain the states of individual tasks \cite{schafer2019user}. Similarly, 
to act on MPI completion events, one needs to regularly test a set of active MPI requests.
The lack of interoperable MPI progress may lead to multiple test-yield cycles that waste CPU cycles and  cause contention on shared MPI resources.

% The need to extend MPI from user space
As HPC enters the exascale era, MPI faces performance challenges on increasingly hybrid node architectures.
Achieving high performance with MPI implementations is becoming more complex and challenging compared to hand-tuned, non-portable solutions.
%For example, optimal collective algorithms for specific systems may involve strategies such as chunking data for pipelining, strategic caching to avoid overhead in hot loops, and offloading work to accelerators.
%Expecting MPI implementations to be optimized for all system configurations is increasingly unrealistic, and it may require new MPI standards to provide the necessary interfaces for adapting MPI to new systems.
%Consequently, the MPI standard is rapidly growing more complex and still struggling to keep pace with system evolution.
%Addressing these challenges requires collaboration from the entire HPC community.
Researchers need the ability to prototype MPI algorithms and MPI extensions independently of MPI implementations.
The ROMIO project, which prototyped and implemented MPI-IO during MPI-2 standardization~\cite{thakur1999implementing}, is a good example of such an approach.
Today, a potential area for similar innovation is MPI collective operations. An algorithm for a collective operation often involves a collection of communication patterns tied together by a progression schedule. An optimized collective algorithm may integrate both MPI communications and asynchronous local offloading steps, tailored to specific system configurations and application needs. Therefore, exposing and making MPI progress interoperable with user-layer asynchronous tasks will stimulate broader community research activities, driving future advancements in MPI. We aim to address a common debate in MPI standardization meetings -- whether a proposed feature needs to be \textit{in MPI} or whether it can be a library \textit{on top of MPI}. Exposing new progress APIs will facilitate more tightly-coupled libraries, so more features can be built on top of MPI first, rather than directly added into MPI before widespread adoption.

% This work
In this work, we introduce a set of MPI extensions to allow applications to explicitly invoke MPI progress without tying to specific communication calls, thereby enabling applications to manage MPI progress without the complexity of handling individual MPI request objects.
Our previously introduced MPIX Stream concept \cite{mpix-stream} is used to target progress to specific contexts, avoiding multithreading contention issues associated with traditional global progress.
We also recognize that MPI progress can be extended to collate progress for asynchronous tasks in general, simplifying the management of multiple progress mechanisms and avoiding wasting cycles on maintaining multiple progress engines.

\section{Anatomy of Asynchronous Tasks and Role of Progress} \label{progress}
Before discussing how to manage progress, we need to define what is progress and understand its role.

\subsection{MPI's Message Modes} \label{prog-mpi}
To understand the role of progress, we examine how an MPI implementation might send and receive messages.
The following discussion is based on MPICH, but we believe it is applicable to other MPI implementations as well.

Figure~\ref{fig:send_mode} illustrates various modes of \texttt{MPI\_Send} and \texttt{MPI\_Recv}.
When sending a small message, the implementation may immediately copy the message to the Network Interface Card (NIC)\footnote{Here ``NIC'' loosely refers to either hardware operations or software emulations.} and return, marking the send operation as complete (see Figure~\ref{fig:send_mode}(a)). While the actual transmission may still be in progress, the send buffer is safe for the application to use. In MPICH, this is called a lightweight send, which notably does not involve any \wait blocks.

For larger messages, buffering costs can be significant. Instead, the message buffer pointer is passed to the NIC, which transmits the message directly from the buffer. \texttt{MPI\_Send} must wait until the NIC signals completion, as the message buffer remains in use until then. This method, known as eager send mode, involves a single \wait block (see Figure~\ref{fig:send_mode}(b)).

When messages are even larger, early arrival at the receiver can cause issues, such as blocking the receiver's message queue or necessitating temporary buffer copying. To prevent unexpectedly receiving large messages, a handshake protocol is used: the sender sends a Ready to Send (RTS) message and waits for the receiver to post a matching buffer and reply with a Clear to Send (CTS) message. The sender then proceeds to send the message data similarly to eager mode. This rendezvous mode involves two \wait blocks (see Figure~\ref{fig:send_mode}(c)).

\texttt{MPI\_Recv} operations vary as shown in Figure~\ref{fig:send_mode}(d-f). Receiving an eager message, including those sent via lightweight send, involves a single \wait block regardless of whether the message arrives before or after \texttt{MPI\_Recv}. Receiving a rendezvous message requires two \wait blocks.

Additional message modes with more complex protocols, such as pipeline mode, may involve multiple \wait blocks. In pipeline mode, a large message is divided into chunks, and the implementation may control the number of concurrent chunks, leading to an indeterminate number of \wait blocks.

\newcommand{\blockwait}[3]{\node[wait, minimum height=#3 cm] at (#1, #2) {\color{white} \large wait};}
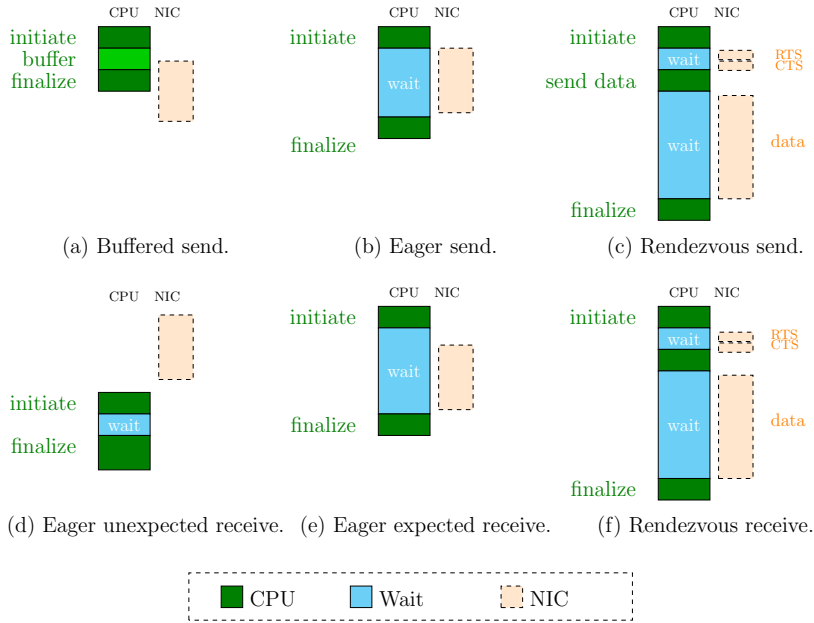
\begin{figure}[t]

\resizebox{0.9\textwidth}{!}{
\begin{tikzpicture}
%  \node[textc, draw] at (2.5, 0) {MPI\_Send};
%  \node[textc, draw] at (8, 0) {MPI\_Recv};
\def\Xa{0}
\def\Xb{6.5}
\def\Xc{13}
\def\Ya{0}
\def\Yb{-6.5}
\def\Yc{-13}

  \node[textc] at (2.5 + \Xa, \Ya -6.5) {\Large (a) Buffered send.};
  \node[textc] at (2.5 + \Xb, \Ya -6.5) {\Large (b) Eager send.};
  \node[textc] at (2.5 + \Xc, \Ya -6.5) {\Large (c) Rendezvous send.};
  \node[textc] at (2.5 + \Xa, \Yb -6.5) {\Large (d) Eager unexpected receive.};
  \node[textc] at (2.5 + \Xb, \Yb -6.5) {\Large (e) Eager expected receive.};
  \node[textc] at (2.5 + \Xc, \Yb -6.5) {\Large (f) Rendezvous receive.};
  \begin{scope}[shift={(5, 0)}]
  \legendA{-14}{Wait}
  \end{scope}

  % buffered send
  \begin{scope}[shift={(\Xa, \Ya)}]
    \node[textc] at (2, -0.9) {CPU};
    \node[textc] at (3, -0.9) {NIC};
    \node[textl] at (1, -1.25) {\Large initiate};
    \node[textl] at (1, -2.25) {\Large finalize};
    \node[textl] at (1, -1.75) {\Large buffer};
    % \node[textr] at (3.75, -2) {transmit};
    \node[cpu, minimum height=0.5cm] at (2, -1) {};
    \node[other, minimum height=0.5cm] at (2, -1.5) {};
    \node[cpu, minimum height=0.5cm] at (2, -2) {};
    \node[nic, minimum height=1.4cm] at (3.2, -1.8) {};
  \end{scope}

  % eager send
  \begin{scope}[shift={(\Xb, \Ya)}]  
    \node[textc] at (2, -0.9) {CPU};
    \node[textc] at (3, -0.9) {NIC};
    \node[textl] at (1, -1.25) {\Large initiate};
    \node[textl] at (1, -3.75) {\Large finalize};
    \node[cpu, minimum height=0.5cm] at (2, -1) {};
    \blockwait{2}{-1.5}{1.6}
    \node[cpu, minimum height=0.5cm] at (2, -3.1) {};
    \node[nic, minimum height=1.5cm] at (3.2, -1.5) {};
  \end{scope}

  % rendezvous send
  \begin{scope}[shift={(\Xc, \Ya)}]  
    \node[textc] at (2, -0.9) {CPU};
    \node[textc] at (3, -0.9) {NIC};
    \node[textl] at (1, -1.25) {\Large initiate};
    \node[textl] at (1, -2.25) {\Large send data};
    \node[textl] at (1, -5.25) {\Large finalize};
    \node[cpu, minimum height=0.5cm] at (2, -1) {};
    \blockwait{2}{-1.5}{0.5}
    \node[cpu, minimum height=0.5cm] at (2, -2) {};
    \blockwait{2}{-2.5}{2.5}
    \node[cpu, minimum height=0.5cm] at (2, -5) {};
    \node[textr] at (4, -1.65) {RTS};
    \node[textr] at (4, -1.9) {CTS};
    \node[textr] at (3.9, -3.65) {\large data};    
    \node[nic, minimum height=0.2cm] at (3.2, -1.55) {};
    \node[nic, minimum height=0.2cm] at (3.2, -1.8) {};
    \node[nic, minimum height=2.4cm] at (3.2, -2.6) {};
  \end{scope}

  % unexpected receive
  \begin{scope}[shift={(\Xa, \Yb - 2)}]
    \node[textc] at (2, 1.0) {CPU};
    \node[textc] at (3, 1.0) {NIC};
    \node[textl] at (1, -1.25) {\Large initiate};
    \node[textl] at (1, -2.25) {\Large finalize};
    \node[cpu, minimum height=0.5cm] at (2, -1) {};
    \blockwait{2}{-1.5}{0.5}
    \node[cpu, minimum height=0.8cm] at (2, -2) {};
    \node[nic, minimum height=1.5cm] at (3.2, 0.8) {};
  \end{scope}

  % expected receive
  \begin{scope}[shift={(\Xb, \Yb)}]
    \node[textc] at (2, -0.9) {CPU};
    \node[textc] at (3, -0.9) {NIC};
    \node[textl] at (1, -1.25) {\Large initiate};
    \node[textl] at (1, -3.75) {\Large finalize};
    \node[cpu, minimum height=0.5cm] at (2, -1) {};
    \blockwait{2}{-1.5}{2.0}
    \node[cpu, minimum height=0.5cm] at (2, -3.5) {};
    \node[nic, minimum height=1.5cm] at (3.2, -1.9) {};
  \end{scope}

  % rendezvous receive
  \begin{scope}[shift={(\Xc, \Yb)}]
    \node[textc] at (2, -0.9) {CPU};
    \node[textc] at (3, -0.9) {NIC};
    \node[textl] at (1, -1.25) {\Large initiate};
    \node[textl] at (1, -5.25) {\Large finalize};
    \node[cpu, minimum height=0.5cm] at (2, -1) {};
    \blockwait{2}{-1.5}{0.5}
    \node[cpu, minimum height=0.5cm] at (2, -2) {};
    \blockwait{2}{-2.5}{2.5}
    \node[cpu, minimum height=0.5cm] at (2, -5) {};
    \node[textr] at (3.9, -1.65) {RTS};
    \node[textr] at (3.9, -1.9) {CTS};
    \node[textr] at (3.9, -3.65) {\large data};
    \node[nic, minimum height=0.2cm] at (3.2, -1.6) {};
    \node[nic, minimum height=0.2cm] at (3.2, -1.85) {};
    \node[nic, minimum height=2.4cm] at (3.2, -2.6) {};
  \end{scope}
\end{tikzpicture}
}
  \caption{Common communication modes: (a) Buffered eager send; (b) Normal eager send;  (c) Rendezvous send; (d) Receiving an eager message that arrived before posting the receive; (e) Receiving an eager message that arrived after posting the receive; (f) Receiving a rendezvous message.}
  \label{fig:send_mode}
\end{figure}

\subsection{Nonblocking and Asynchronous Task Patterns} \label{prog-patterns}
The \wait blocks in Figure~\ref{fig:send_mode} illustrate why \texttt{MPI\_Send} and \texttt{MPI\_Recv} are considered blocking operations. By focusing on the \wait blocks in the block diagram, we can abstract tasks such as \texttt{MPI\_Send} and \texttt{MPI\_Recv} into three patterns, as shown in Figure~\ref{fig:task_patterns}: tasks that do not \textit{wait}, tasks that contain a single \wait block, and tasks with multiple \wait blocks. During a \wait block, the task is executed on a hardware device such as a NIC or GPU, within the OS kernel, or within a separate execution context such as a thread or process.

\begin{figure}[t]
    \centering
    \begin{minipage}[t]{0.45\textwidth}
    \centering
    \resizebox{\textwidth}{!}{
\begin{tikzpicture}

\begin{scope}[shift={(0, -1)}] 
\node[cpu, minimum height=1cm] at (0, 0) {};

\node[cpu, minimum height=0.3cm] at (3, 0) {};
\node[wait, minimum height=1.5cm] at (3, -0.3) {};
\node[cpu, minimum height=0.3cm] at (3, -1.8) {};
\node[nic, minimum height=1.5cm] at (4.1, -0.3) {};

\node[cpu, minimum height=0.3cm] at (6, 0) {};
\node[wait, minimum height=0.3cm] at (6, -0.3) {};
\node[cpu, minimum height=0.3cm] at (6, -0.6) {};
\draw[wait] (5.4, -0.9) -- (6.6, -0.9) -- (6.6, -1.3) ..controls (6.3, -1.6) and (5.7, -1.0) .. (5.4, -1.3) -- cycle;
\draw[wait] (5.4, -1.9) -- (6.6, -1.9) -- (6.6, -1.5) ..controls (6.3, -1.8) and (5.7, -1.2) .. (5.4, -1.5) -- cycle;
%\node[wait, minimum height=1.0cm] at (6, -0.9) {};
\node[cpu, minimum height=0.3cm] at (6, -1.9) {};
\node[wait, minimum height=0.3cm] at (6, -2.2) {};
\node[cpu, minimum height=0.3cm] at (6, -2.5) {};
\node[nic, minimum height=1.5cm] at (7.1, -0.9) {};
\end{scope}

\node[textc] at (3.2, 0) {\LARGE Asynchronous Task Patterns};
\subabc{-6.7}
\legendA{-7.5}{Wait}
\end{tikzpicture}
}
    \caption{Task patterns: (a) A task with no blocking parts; (b) A task with a single blocking part; (c) A task with multiple blocking parts.}
    \label{fig:task_patterns}
    \end{minipage}\qquad
    \begin{minipage}[t]{0.45\textwidth}
    \centering
    \resizebox{\textwidth}{!}{
\begin{tikzpicture}
  \node[textc] at (3.2, 0) {\LARGE Nonblocking Start};
  \node[textc] at (3.2, -2.5) {\LARGE Completion};

\begin{scope}[shift={(0, -0.5)}] 
\node[cpu, minimum height=1.0cm] at (0, 0) {};
\node[cpu, minimum height=0.3cm] at (3, 0) {};
\node[cpu, minimum height=0.3cm] at (6, 0) {};
\end{scope}

\begin{scope}[shift={(0, -2)}]    
\node[cpu, minimum height=0.2cm] at (0, -1.3) {};

\node[wait, minimum height=1.5cm] at (3, -1.3) {};
\node[cpu, minimum height=0.3cm] at (3, -2.8) {};
\node[nic, minimum height=1.5cm] at (4.1, -1.3) {};

\node[wait, minimum height=0.3cm] at (6, -1.3) {};
\node[cpu, minimum height=0.3cm] at (6, -1.6) {};
\draw[wait] (5.4, -1.9) -- (6.6, -1.9) -- (6.6, -2.3) ..controls (6.3, -2.6) and (5.7, -2.0) .. (5.4, -2.3) -- cycle;
\draw[wait] (5.4, -2.9) -- (6.6, -2.9) -- (6.6, -2.5) ..controls (6.3, -2.8) and (5.7, -2.2) .. (5.4, -2.5) -- cycle;
\node[cpu, minimum height=0.3cm] at (6, -2.9) {};
\node[wait, minimum height=0.3cm] at (6, -3.2) {};
\node[cpu, minimum height=0.3cm] at (6, -3.5) {};
\node[nic, minimum height=1.5cm] at (7.1, -1.9) {};
\end{scope}

\subabc{-6.7}
\legendA{-7.5}{Wait}
\end{tikzpicture}
}
    \caption{Nonblocking tasks: (a) A task with no blocking parts; (b) A task with a single blocking part; (c) A task with multiple blocking parts.}
    \label{fig:nonblocking}
    \end{minipage}%
    
\end{figure}

\begin{figure}
    \centering
    \begin{minipage}[t]{0.45\textwidth}
      \centering
      \resizebox{\textwidth}{!}{
\begin{tikzpicture}
%  \node[textc, draw] at (2.5, 0) {MPI\_Send};
%  \node[textc, draw] at (8, 0) {MPI\_Recv};

\node[cpu, minimum height=1.0cm] at (0, 0) {};
\node[comp, minimum height=1.0cm] at (0, -1.0) {};
\node[cpu, minimum height=0.2cm] at (0, -2.0) {};

\node[cpu, minimum height=0.3cm] at (3, 0) {};
\node[comp, minimum height=1.0cm] at (3, -0.3) {};
\node[wait, minimum height=0.5cm] at (3, -1.3) {};
\node[cpu, minimum height=0.3cm] at (3, -1.8) {};
\node[nic, minimum height=1.5cm] at (4.1, -0.3) {};

\node[cpu, minimum height=0.3cm] at (6, 0) {};
\node[comp, minimum height=1.0cm] at (6, -0.3) {};
\node[wait, minimum height=0.3cm] at (6, -1.3) {};
\node[cpu, minimum height=0.3cm] at (6, -1.6) {};
\draw[wait] (5.4, -1.9) -- (6.6, -1.9) -- (6.6, -2.3) ..controls (6.3, -2.6) and (5.7, -2.0) .. (5.4, -2.3) -- cycle;
\draw[wait] (5.4, -2.9) -- (6.6, -2.9) -- (6.6, -2.5) ..controls (6.3, -2.8) and (5.7, -2.2) .. (5.4, -2.5) -- cycle;
\node[cpu, minimum height=0.3cm] at (6, -2.9) {};
\node[wait, minimum height=0.3cm] at (6, -3.2) {};
\node[cpu, minimum height=0.3cm] at (6, -3.5) {};
\node[nic, minimum height=1.5cm] at (7.1, -1.9) {};

\node[textc] at (3.2, 1) {\LARGE Computation/Communicaton Overlap};
\subabc{-6}
\legend{-7}{Wait}
\end{tikzpicture}
}
      \caption{Computation/communication overlap: (a) Communication with no blocking parts; (b) Communication with single blocking part; (c) Communication with multiple blocking parts.}
      \label{fig:overlap}
    \end{minipage}\qquad
    \begin{minipage}[t]{0.45\textwidth}
      \resizebox{\textwidth}{!}{
\begin{tikzpicture}
\begin{scope}[shift={(0, 0)}] 
\node[cpu, minimum height=0.8cm] at (0, 0) {};
\node[comp, minimum height=0.35cm] at (0, -0.8) {};
\node[wait, minimum height=0.2cm] at (0, -1.15) {};
\node[comp, minimum height=0.35cm] at (0, -1.35) {};
\node[wait, minimum height=0.2cm] at (0, -1.7) {};
\draw[comp] (-0.6, -1.9) -- (0.6, -1.9) -- (0.6, -2.3) ..controls (0.3, -2.6) and (-0.3, -2.0) .. (-0.6, -2.3) -- cycle;
\draw[comp] (-0.6, -2.9) -- (0.6, -2.9) -- (0.6, -2.5) ..controls (0.3, -2.8) and (-0.3, -2.2) .. (-0.6, -2.5) -- cycle;
\node[wait, minimum height=0.2cm] at (0, -2.9) {};
\node[comp, minimum height=0.35cm] at (0, -3.1) {};
\node[wait, minimum height=0.2cm] at (0, -3.45) {};
\node[comp, minimum height=0.35cm] at (0, -3.65) {};
\node[wait, minimum height=0.1cm] at (0, -4.0) {};
\node[cpu, minimum height=0.8cm] at (0, -4.1) {};

\node[nic, minimum height=1.2cm] at (1.3, -0.8) {};
\node[nic, minimum height=1.0cm] at (1.3, -3.2) {};

\end{scope}

\begin{scope}[shift={(5, 0)}]    
\node[cpu, minimum height=0.8cm] at (0, 0) {};
\node[comp, minimum height=2.8cm] at (0, -0.8) {};
\node[wait, minimum height=0.3cm] at (0, -3.6) {};
\node[cpu, minimum height=0.8cm] at (0, -3.9) {};

\foreach \i in {0,...,12} {
  \node[wait, minimum height=0.25cm] at (1.6, -\i*0.4+0.1) {};
}

\node[nic, minimum height=1.2cm] at (2.9, -0.8) {};
\node[nic, minimum height=1.0cm] at (2.9, -2.9) {};

\end{scope}

\node[textc] at (3.2, 1) {\LARGE Progress Schemes};
\subab{-6}
\legend{-7}{Test}
\end{tikzpicture}
}
      \caption{Remedies for the lack of progress: (a) Intersperse progress tests inside computations; (b) Use a dedicated thread to continuously poll progress.}
      \label{fig:prog_thread}
    \end{minipage}
\end{figure}

The \wait block is often implemented as a busy poll loop, which wastes CPU cycles while the offloaded task is still in progress. Conversely, if the offloaded task finishes and the completion event is not immediately polled and acted upon, it can delay subsequent dependent work, adding latency to the workflow.

Rather than immediately waiting for an asynchronous task to complete, a program can, in principle, perform other jobs that do not depend on the pending task. This is the idea behind MPI's nonblocking APIs. A nonblocking operation splits a corresponding blocking operation into two parts: starting and completion. For example, \texttt{MPI\_Send} is divided into \texttt{MPI\_Isend} and \texttt{MPI\_Wait}.

Figure~\ref{fig:nonblocking} illustrates how blocking patterns in Figure~\ref{fig:task_patterns} are split into nonblocking patterns. If the task does not contain any \wait blocks (Figure~\ref{fig:nonblocking}(a)), the split into a nonblocking pattern is somewhat arbitrary, but typically the entire operation is completed in the starting call, and the completion call will return immediately. If the task contains a single \wait block (Figure~\ref{fig:nonblocking}(b)), it is naturally split just before the \wait block. For tasks with multiple \wait blocks (Figure~\ref{fig:nonblocking}(c)), the split occurs before the first \textit{wait}. Generally, the starting call should avoid any \wait blocks to preserve the nonblocking semantics.

Viewing MPI operations through the lens of \wait patterns generalizes MPI nonblocking operations to common asynchronous programming patterns. For instance, the async/await syntax \cite{okur2014study} in some programming languages provides a concise method to describe the \wait patterns in a task.
Event-driven programming \cite{haller2006event}, on the other hand, expresses the code following the \wait block as event callbacks.
In MPI, these async patterns are opaque, making MPI progress management obscure.

\subsection{Computation/Communication Overlap} \label{prog-overlap}
One of the primary goals of using nonblocking MPI operations is to achieve overlap between computation and communication. Ideally, the CPU cycles spent in a wait loop should instead be used for computation, enhancing overall efficiency. However, achieving this overlap with MPI is not straightforward.

The concept of computation/communication overlap is illustrated in Figure~\ref{fig:overlap}. Immediately after initiating a nonblocking operation, the program enters a computation phase while the message data transmission is handled by the NIC hardware or another offloading device. Once the computation phase completes, the program resumes the \wait for the nonblocking operation. If the communication has finished by then, the final \wait returns immediately; otherwise, the \wait time is significantly reduced. This overlap maximizes efficiency, improving overall performance and reducing time to solution.

The ideal overlap can be easily achieved for the case in Figure~\ref{fig:overlap}(b), where a single \wait in the nonblocking operation allows for effective overlap. In contrast, Figure~\ref{fig:overlap}(a) shows a scenario with no \wait block to save, offering no additional overlap compared to the blocking case. Converting a blocking operation without a \wait into a nonblocking one only introduces overhead due to the creating and finalizing of a task handle (i.e., an MPI request). However, in most MPI implementations, this overhead is negligible.
The situation is more complex in Figure~\ref{fig:overlap}(c), where multiple \wait blocks are present, and computation only overlaps with the first \wait block. This initial overlap is often insignificant compared to the total combined \wait time. For instance, in a simple rendezvous message, the first \wait involves waiting for a small protocol handshake, while the bulk of the message transmission occurs during the second \textit{wait}. As a result, the opportunity for significant overlap is missed in such cases.

\subsection{Role of Progress}
The key issue with the scenario depicted in Figure~\ref{fig:overlap}(c) is the lack of progress. After the first \wait ends, a small block of code needs to run to initiate the second \textit{wait}.
For asynchronous tasks with multiple \wait blocks, this small block of code after each \wait block must run to trigger the subsequent asynchronous tasks for the next \textit{wait}.
Polling for completion events and running the handlers to initiate the following asynchronous tasks constitute progress.
Without adequate progress, the next steps in the asynchronous task are delayed, resulting in degraded performance.

There are two remedies for this lack of progress. One is to intersperse \texttt{MPI\_Test} calls within the computation, as illustrated in Figure~\ref{fig:prog_thread}(a). However, this approach has at least three drawbacks. First, breaking the computation into parts and interspersing it with progress calls significantly increases code complexity and may not always be feasible. For example, the computation might be encapsulated in an opaque function, or the bulk of it might be spent in a math routine from an external library. Second, if progress polling is too frequent, many polls will waste CPU cycles without benefit, and the context switching between computation and progress polling adds overhead. Consequently, frequent polling decreases performance. Third, if progress polling is too sparse, the likelihood of polling just after a communication step completes is low, resulting in imperfect computation/communication overlap.

An alternative solution is to use a dedicated CPU thread for polling progress. This approach ensures sufficient progress and maximizes the overlap between computation and communication. However, it also wastes CPU cycles when a communication step is not ready and occupies an entire CPU core. While modern HPC systems often have many cores, dedicating a CPU core for progress can be acceptable for some applications. However, this becomes problematic when multiple processes are launched on a single node. If each process has its own progress thread, it can quickly exhaust CPU cores and severely impact performance. Additionally, applications may use other asynchronous subsystems besides MPI, each potentially requiring its own progress thread, leading to competition for limited core resources.

\subsection{Managing MPI Progress}
Implementing a progress thread with MPI is challenging, primarily because MPI does not provide explicit APIs for invoking MPI progress. MPI progress is largely hidden within the implementation, with the assumption that a system-optimized MPI will provide strong progress, thereby reducing the need for explicit progress. However, as discussed earlier, this assumption may not always hold true.
Calling any MPI function may or may not invoke MPI progress, and when it does, it may not serve global progress. For instance, calling \texttt{MPI\_Test} on one MPI request may not necessarily advance other MPI requests.
Furthermore, any MPI function that invokes progress may contend for locks with another thread calling MPI functions.
Managing MPI progress can feel almost magical when it works, but extremely frustrating when it fails.

One of the more explicit ways to invoke MPI progress is by calling \texttt{MPI\_Test} (or any of its variants). \texttt{MPI\_Test} requires an explicit MPI request parameter. However, a dedicated progress thread is often isolated from the context that initiates the actual MPI operations, making it challenging to synchronize MPI requests between the computation threads and the progress thread. This synchronization of MPI request objects imposes a significant burden on a many-task system design.
Therefore, to enable applications to build effective progress engines, MPI needs to provide mechanisms for invoking progress independent of specific MPI requests.

\subsection{Collating Progress} \label{prog-collat}
In addition to the inconvenience of having to use an MPI request to call \texttt{MPI\_Test}, polling progress for each individual MPI requests is also inefficient.
It is more efficient to poll for all events,  process them one by one, and then check whether a specific MPI request has been completed from the event handling.
This is referred to as collating progress.
Collating progress ensures that all parts of the program are progressing towards completion, rather than waiting unnecessarily for each individual operation to complete sequentially. Collating progress can help improve overall application performance by reducing bottlenecks and enhancing concurrency.

In addition to collating progress for network operations, an MPI library internally needs to manage the progress of multiple asynchronous subsystems. For example, data transfer may involve GPU device memory, meaning a conventional MPI send and receive could include asynchronous memory copy operations between host and device memory. MPI-IO may introduce asynchronous storage I/O operations. Collectives are often implemented as a series of nonblocking point-to-point communications following a multi-stage pattern similar to Figure~\ref{fig:nonblocking}(c). Additionally, MPI communication may internally utilize different subsystems depending on whether the communication is between on-node processes or inter-node processes.
All these asynchronous subsystems require progress, and it is often more convenient and efficient to collate them.

Listing~\ref{lst:mpich-progress} shows the pseudocode of MPICH's internal progress function.

\begin{lstlisting}[language=C, frame=tb, caption=MPICH's progress function, label=lst:mpich-progress]
int MPIDI_progress_test(MPID_Progress_state *state) {
    int mpi_errno = MPI_SUCCESS, made_progress = 0;

    /* asynchronous datatype pack/unpack */
    Datatype_engine_progress(&made_progress);
    if (made_progress) goto fn_exit;

    /* collective algorithms */
    Collective_sched_progress(&made_progress);
    if (made_progress) goto fn_exit;

    /* intranode shared memory communication */
    Shmem_progress(&made_progress);
    if (made_progress) goto fn_exit;

    /* internode netmod communication */
    Netmod_progress(&made_progress);
    if (made_progress) goto fn_exit;

  fn_exit:
    return mpi_errno;
}
\end{lstlisting}

This progress routine is called whenever an MPI function requires progress. Collating progress assumes that the cost is negligible if a subsystem has no pending tasks. For the datatype engine, collective, and shared memory (shmem) subsystems, an empty poll incurs a cost equivalent to reading an atomic variable. However, this is not always the case with netmod progress, so we place netmod progress last and skip it whenever progress is made with other subsystems.
Additionally, MPICH's progress function accepts a state variable from the calling stack, providing an opportunity for the caller to tune the progress performance according to the context.
For example, from a context where only netmod progress is needed, the progress state can be set to skip progress for all other subsystems.
Since MPI implementations already perform collated event-based progress internally, exposing MPI progress as an explicit API for applications is straightforward.

\subsection{Case for Interoperable MPI Progress} \label{prog-event}
As we have discussed, the patterns of MPI internal operations are similar to those of general asynchronous tasks that applications may create. Therefore, the design and optimization of MPI progress should be applicable to application-layer asynchronous tasks as well. In fact, current MPI implementations already handle several async subsystems internally, making it straightforward to extend this capability to work with external tasks. We refer to this concept as ``interoperable MPI progress.''

Interoperable MPI progress provides applications with a mature progress engine, eliminating the need to create and maintain separate progression mechanisms for each new async system. Additionally, integrating user-layer progress within MPI progress is more convenient and efficient.

Another advantage of interoperable MPI progress is that it allows for the implementation and extension of MPI subsystems at the user level. For instance, users could implement collectives in user space by adding a progress hook into MPI's progress, similar to the \texttt{Collective\_sched\_progress} in Listing~\ref{lst:mpich-progress}. This approach promotes a modular design where parts of MPI are built on top of a core MPI implementation, enhancing both flexibility and stability. Furthermore, a core MPI set that facilitates the building of MPI extensions can stimulate broader community research activities and infuse new life into MPI.

% Explain the rationale
% Explain the API
% Discuss the concerns

\section{MPICH Extensions to Enable Progress and Event-Driven Programming} \label{extensions}
In this section, we present new extension APIs developed in MPICH that enable applications to more effectively manage MPI progress and to extend MPI through interoperable MPI progress.

\subsection{MPIX Streams} % allowing contention-free concurrent progress
First, to address the issue of lacking execution context in MPI operations and MPI progress, we utilize the MPIX Stream concept proposed in our previous work\cite{mpix-stream}. An MPIX Stream represents an internal communication context within the MPI library, defined as a serial execution context. All operations attached to an MPIX Stream are required to be issued in a strict serial order, eliminating the need for lock protection within the MPI library. This allows a multithreaded application to achieve maximum parallel performance without the penalty of lock contention.
Additionally, MPIX Streams can be applied to progress, targeting operations in a specific stream.

An MPIX Stream is created using the following API:
\begin{lstlisting}[style=proto]
int MPIX_Stream_create(MPI_Info info, MPIX_Stream *stream)
\end{lstlisting}

To use an MPIX Stream in MPI communications, you must first create a stream communicator with the following function:
\begin{lstlisting}[style=proto]
int MPIX_Stream_comm_create(MPI_Comm parent_comm, MPIX_Stream stream, MPI_Comm *stream_comm)
\end{lstlisting}

A stream communicator can be used the same way as a conventional MPI communicator, except that all operations on a stream communicator will be associated with the corresponding MPIX Stream context.
While an MPIX Stream is naturally suited for a thread context, it can also be applied to any semantically serial construct.
For example, the serial context can be manually enforced through thread barriers, or originate from a specific runtime such as a CUDA stream.
Info hints offer a flexible mechanism for implementations to extend support and apply specific optimizations.
For more detailed information on MPIX Streams, please refer to our previous work\cite{mpix-stream}.

\subsection{Explicit MPI Progress} % allowing progress without managing request handles
To address the need for making MPI progress without being tied to specific MPI requests,
We propose an API that allows applications to advance MPI progress for a specific MPIX Stream:

\begin{lstlisting}[style=proto]
int MPIX_Stream_progress(MPIX_Stream stream)
\end{lstlisting}

%Using an MPIX Stream is optional. 
If context separation is not a concern, the application can use the default stream, \texttt{MPIX\_STREAM\_NULL}.
However, using explicit MPIX Streams can help applications avoid unnecessary thread contention.
Additionally, MPIX Streams can be used to separate progress for different subsystems, especially if some subsystems are sensitive to latency and need to avoid collating progress.
For example, in Listing~\ref{lst:mpich-progress}, hints can be provided to the MPIX Streams to skip \texttt{Netmod\_progress} if the subsystem does not depend on inter-node communication.

\subsection{MPIX Async Extension} % allowing collating user-layer progress with MPI progress
\texttt{MPIX\_Stream\_progress} allows applications to incorporate MPI progress into their progression schemes.
However, to make MPI progress truly interoperable, we also need a mechanism for applications to add progress hooks into the MPI progress system.
This is accomplished with the following extension:
\begin{lstlisting}[style=proto]
int MPIX_Async_start(MPIX_Async_poll_function poll_fn, void *extra_state, MPIX_Stream stream)
\end{lstlisting}

The \texttt{poll\_fn} parameter is a user-defined progress hook function that is called from within MPI progress (e.g., inside \texttt{MPIX\_Stream\_progress} or \texttt{MPI\_Test}) along with MPI's internal progress hooks (See Listing~\ref{lst:mpich-progress}).
\texttt{extra\_state} is a user-defined handle or a state pointer that will be passed back to \texttt{poll\_fn}. 
The \texttt{stream} parameter attaches the task to the corresponding MPIX Stream, including the default stream, \texttt{MPIX\_STREAM\_NULL}. 
\texttt{poll\_fn} has the following signature:
\begin{lstlisting}[style=proto]
typedef struct MPIR_Async_thing *MPIX_Async_thing;
typedef int (MPIX_Async_poll_function)(MPIX_Async_thing);
\end{lstlisting}

An opaque \texttt{struct} pointer, \texttt{MPIX\_Async\_thing}, is used instead of directly passing \texttt{extra\_state} back to \texttt{poll\_fn}.
This provides some flexibility for implementations to support more features.
\texttt{MPIX\_Async\_thing} combines application-side context (i.e., \texttt{extra\_state}) and the implementation-side context.
Inside \texttt{poll\_fn}, the original \texttt{extra\_state} can be readily retrieved with:
\begin{lstlisting}[style=proto]
void *MPIX_Async_get_state(MPIX_Async_thing async_thing)
\end{lstlisting}

One example feature the implementation may support is to allow applications to spawn additional async tasks while progressing a pending task inside \texttt{poll\_fn}.
This is accomplished with the following API:
\begin{lstlisting}[style=proto]
void *MPIX_Async_spawn(MPIX_Async_thing async_thing, MPIX_Async_poll_function poll_fn, void *extra_state, MPIX_Stream stream)
\end{lstlisting}

The newly spawned tasks are temporarily stored inside \texttt{async\_thing} and will be processed after \texttt{poll\_fn} returns.
This allows the implementation to avoid potential recursion and the need for global queue protection before calling \texttt{poll\_fn}.

\texttt{poll\_fn} returns either
%one of the following codes:
%\begin{lstlisting}[style=proto]
%enum {
%    MPIX_ASYNC_PENDING = 0,
%    MPIX_ASYNC_DONE = 1,
%};
%\end{lstlisting}
\texttt{MPIX\_ASYNC\_PENDING} if the async task is in progress or \texttt{MPIX\_ASYNC\_DONE} if the async task is completed.
Before \texttt{poll\_fn} returns \texttt{MPIX\_ASYNC\_DONE}, it must clean up the application context associated with the async task, typically by freeing the structure behind \texttt{extra\_state}. The MPI library will then free the context behind \texttt{MPIX\_Async\_thing}.

The MPIX Async interface allows users to extend MPI's functionality and integrate custom progression schemes into MPI progress.
For example, an MPI collective can be viewed as a fixed task graph composed of individual operations and their dependencies.
By defining \texttt{poll\_fn}, one can advance a specific task graph for a custom collective algorithm within MPI progress.
Integrating into MPI progress simplifies the process by eliminating the need for constructing separate progress mechanisms and avoiding performance issues such as managing MPI request objects, extra progress threads, and thread contention.

\subsection{Completion Query on MPI Requests}
When a task finishes its computation for a given stage, it must wait for the completion of its dependent nonblocking operations.
However, invoking redundant progress in the presence of a progress engine is unnecessary and undesirable.
Instead of using \texttt{MPI\_Wait}, we need a request completion query function that does not trigger progress.
The following API provides this functionality:
\begin{lstlisting}[style=proto]
bool MPIX_Request_is_complete(MPI_Request request)
\end{lstlisting}

The implementation simply queries an atomic flag for the request, resulting in minimal overhead when repeatedly polling this function. Importantly, there are no side effects that would interfere with other requests or other progress calls.
\texttt{MPIX\_Request\_is\_complete} is also useful in the MPIX Async \texttt{poll\_fn} when the application-layer task is built upon MPI operations.
Each MPI progress may internally use a context to coordinate various parts, thus, invoking progress recursively inside the \texttt{poll\_fn} is prohibited.

\subsection{Programming Scheme}
\begin{figure}[t]
    \centering
    \includegraphics[width=0.7\textwidth]{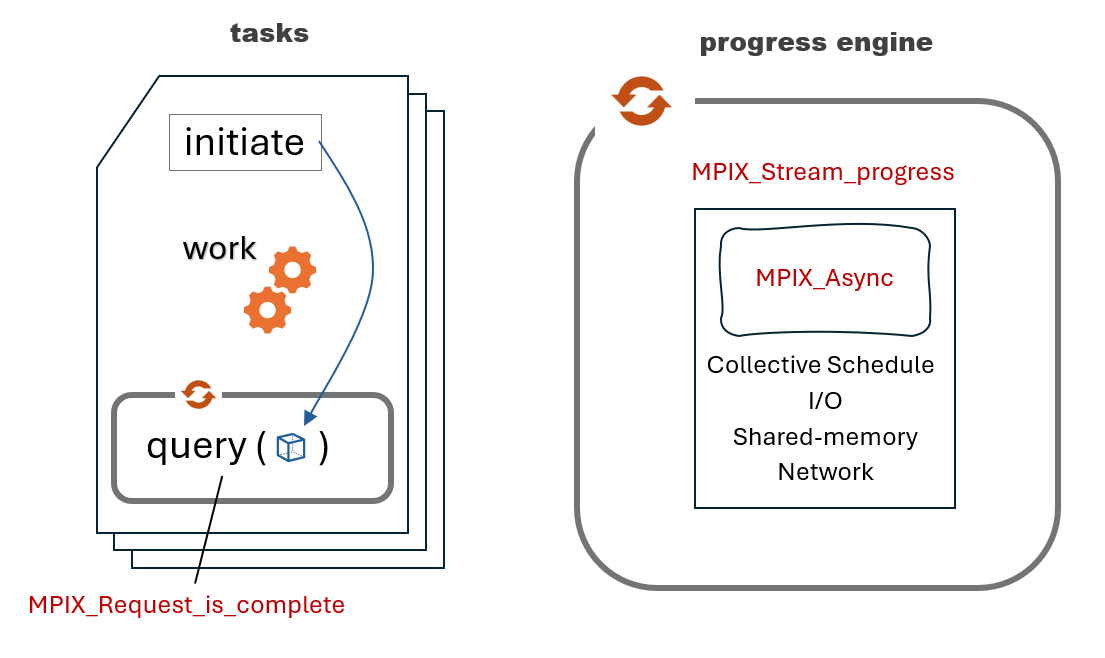}
    \caption{Programming scheme that separates operation context and progress engine.}
    \label{fig:prog-scheme}
\end{figure}

Figure~\ref{fig:prog-scheme} illustrates how the above extension APIs fit together in a programming scheme where the task contexts are separated from the progress engine. (Due to space constraints, we do not provide a code listing.)
In this scheme, programmers utilize asynchronous APIs, such as MPI nonblocking operations, to overlap with computational work. However, to achieve maximum overlap, a progress engine is generally required to ensure that asynchronous operations continue to progress while the main threads are engaged in computation.
By separating individual task contexts, such as MPI request objects, from the progress engine, the design is simplified, and the additional latency that may occur from synchronizing request objects between tasks and the progress engine is avoided.
\texttt{MPIX\_Stream\_progress} enables applications to invoke MPI progress without requiring the use of request objects, thus separating task contexts.
\texttt{MPIX\_Request\_is\_complete} allows tasks to synchronize on asynchronous jobs without invoking MPI progress, ensuring orthogonality between tasks and the progress engine.
Finally, MPIX Async extensions enable custom asynchronous operations to be progressed by the MPI progress, thereby making MPI progress interoperable with non-MPI asynchronous designs.
Utilizing MPI progress to advance all asynchronous operations simplifies the design of a progress engine and makes optimizations, such as using MPIX Streams to avoid accidental contention, generally accessible to the entire application.

\section{Examples} \label{examples}
In this section, we present various examples using the MPIX Async extensions.
An important metric for quantifying progress performance is the progress latency, defined as the average elapsed time between a task's completion and when the user code responds to the event.
We use a dummy task to directly measure progress latency.
Unless otherwise noted, our experiments were conducted on a local workstation with an 8-core i7-7820X CPU.

\subsection{Dummy task}
For most of the following examples, we use a dummy task that completes after a predetermined duration.
This simulates an asynchronous job completed via offloading.
Instead of querying a completion status, we check for the elapsed time.
The code is provided in Listing~\ref{code:dummy}.
By presetting the duration for the dummy task to complete, we can measure the latency of the progress engine's response to the completion event.

\begin{lstlisting}[style=example, caption=An example of dummy async task using MPIX Async extensions, label=code:dummy]
#define TASK_DURATION 1.0
#define NUM_TASKS 10

struct dummy_state {
    double wtime_finish;
};

static int dummy_poll(MPIX_Async_thing thing)
{
    struct dummy_state *p = MPIX_Async_get_state(thing);
    double wtime = MPI_Wtime();
    if (wtime >= p->wtime_finish) {
        /* double diff = (wtime - p->wtime_finish) * 1e6; */    
        free(p);
        return MPIX_ASYNC_DONE;
    }
    return MPIX_ASYNC_NOPROGRESS;
}

static void add_async(void)
{
    struct dummy_state *p = malloc(sizeof(struct dummy_state));
    p->wtime_finish = MPI_Wtime() + TASK_DURATION;
    MPIX_Async_start(dummy_poll, p);
}

int main(int argc, const char **argv)
{
    MPI_Init(NULL, NULL);
    
    for (int i = 0; i < NUM_TASKS; i++) {
        add_async();
    }

    /* In this example, MPI_Finalize will spin progress until all async tasks completes */
    MPI_Finalize();

    return 0;
}
\end{lstlisting}

In Listing~\ref{code:dummy}, there is no active progress management since we only care about the tasks being launched and completed without synchronization.
\texttt{MPI\_Finalize} will always continue progress until all asynchronous tasks are done.
To add synchronization, global states or references to synchronization variables are needed. This is demonstrated in Listing~\ref{code:sync}, where the new code from the previous listing is highlighted.
We also added a wait-loop after adding the tasks. Since there is no computation in our example, we simply drive the progress within the main thread.

\definecolor{mycolor}{rgb}{0.9, 0.9, 0.8}

\begin{lstlisting}[style=example, caption={Adding synchronization counter, wait-progress loop, and stubs for latency benchmark}, label=code:sync, escapechar=@]
#define TASK_DURATION 1.0
#defile NUM_TASKS 10

struct dummy_state {
    double wtime_finish;
    int *counter_ptr;
};

void add_stat(double latency);  /* implementation omitted */
void report_stat(void);         /* implementation omitted */

static int dummy_poll(MPIX_Async_thing thing)
{
    struct dummy_state *p = MPIX_Async_get_state(thing);
    double wtime = MPI_Wtime();
    if (wtime >= p->wtime_finish) {
        add_stat(wtime - p->wtime_finish) * 1e6;
        (*(p->counter_ptr))--;
        free(p);
        return MPIX_ASYNC_DONE;
    }
    return MPIX_ASYNC_NOPROGRESS;
}

static void add_async(int *counter_ptr)
{
    struct dummy_state *p = malloc(sizeof(struct dummy_state));
    p->wtime_finish = MPI_Wtime() + TASK_DURATION;
    p->counter_ptr = counter_ptr;
    MPIX_Async_start(dummy_poll, p);
}

int main(int argc, const char **argv)
{
    MPI_Init(NULL, NULL);
    
    int counter = NUM_TASKS;
    for (int i = 0; i < NUM_TASKS; i++) {
        add_async(&counter);
    }

    /* Essentially a wait block */
    while (counter > 0) {
        MPIX_Stream_progress(MPIX_STREAM_NULL);
    }

    report_stat();
    
    MPI_Finalize();

    return 0;
}
\end{lstlisting} 

\subsection{Performance Factors in Async Progress}
Several performance factors influence the average response latency to event completion in asynchronous progress.

\subsubsection{Number of pending tasks} \label{eg:pending}
\begin{figure}
    \centering
    \begin{minipage}[t]{0.45\textwidth}
        \centering
        \includegraphics[width=0.8\textwidth]{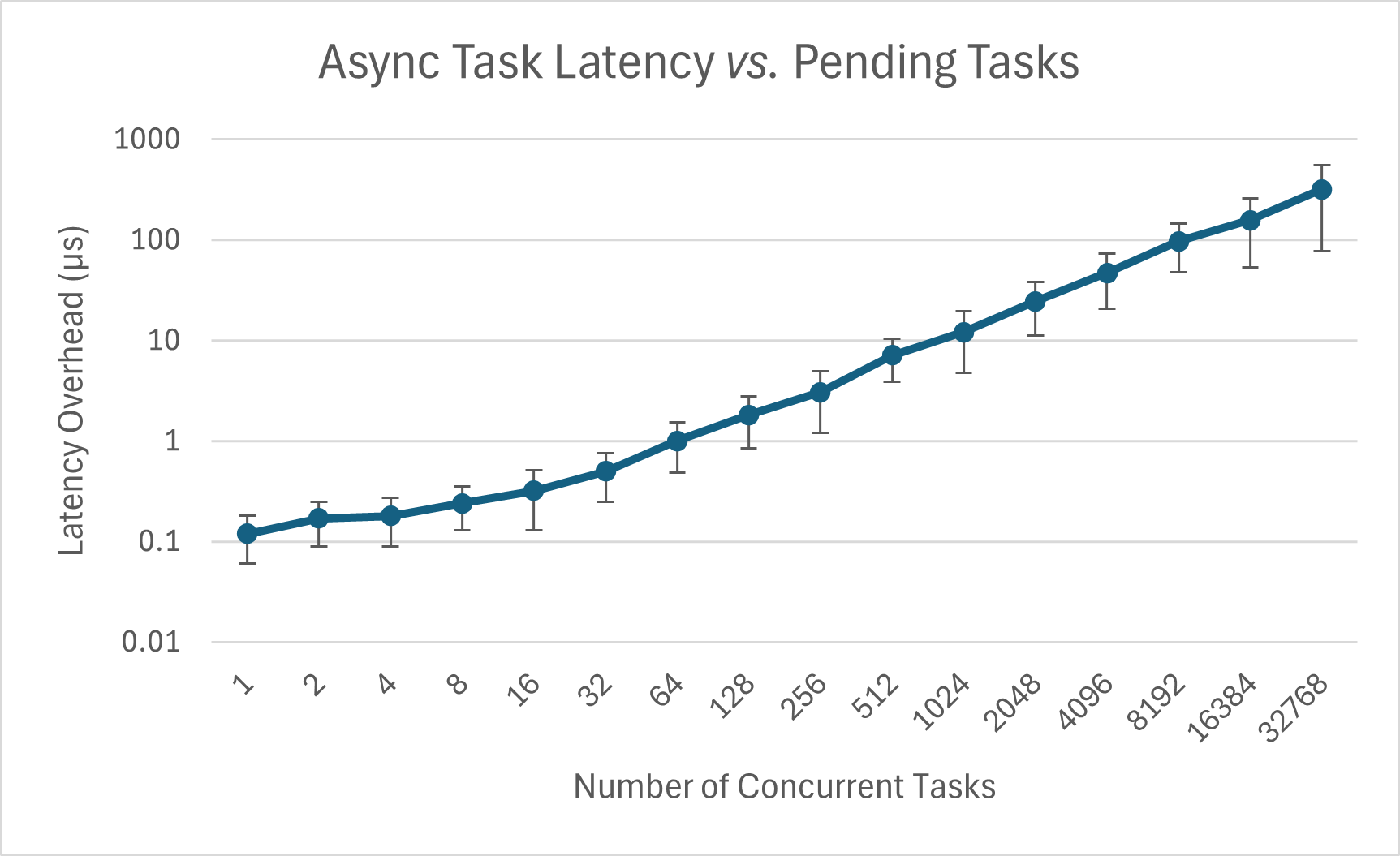}
        \caption{Latency overhead in microseconds as the number of pending async tasks increases.}
        \label{fig:latency}
    \end{minipage}\qquad 
    \begin{minipage}[t]{0.45\textwidth}
        \centering
        \includegraphics[width=0.8\textwidth]{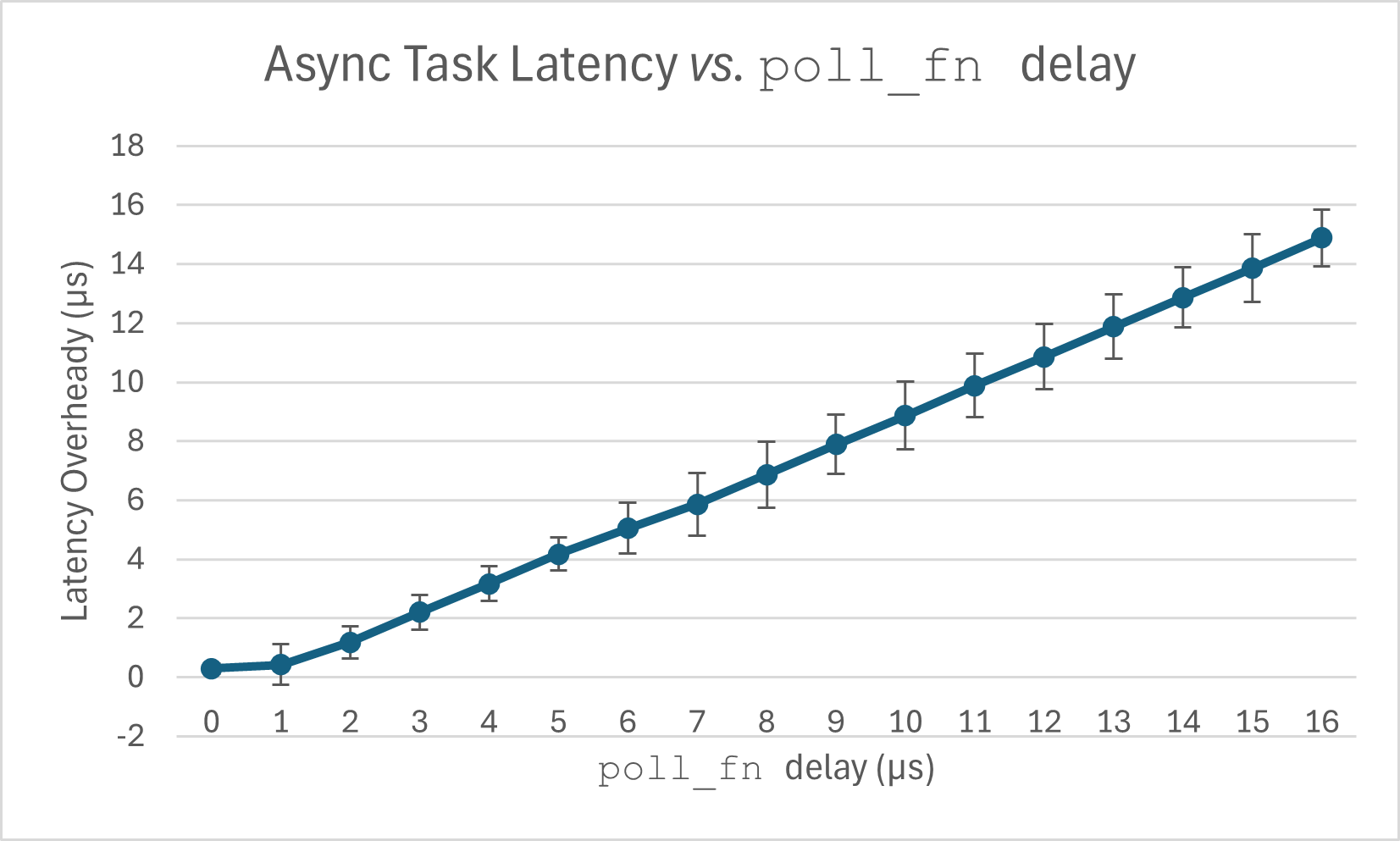}
        \caption{Impact of poll function overhead on event response latency. Each measurement runs 10 concurrent pending tasks. The delay is implemented by busy-polling \texttt{MPI\_Wtime}.}
        \label{fig:poll-overhead}
    \end{minipage}
\end{figure}

During each MPI progress call (e.g., \texttt{MPIX\_Stream\_progress}), MPI will invoke the async \pollfn\ for each pending async task sequentially. The cycles spent processing numerous tasks may delay the response time to a specific task's completion event. Therefore, as the number of pending tasks increases, we expect an increase in response latency. This expectation is confirmed by the experimental results shown in Figure~\ref{fig:latency}.

If all the pending tasks are independent, each progress call must invoke \pollfn\ for every pending task, leading to a performance degradation as the number of pending tasks rises.
Notably, when there are fewer than 32 pending tasks, the latency overhead remains below 0.5 microseconds.

Most applications do not create thousands of independent tasks randomly. Typically, tasks have dependencies on each other, forming a task graph, or they are grouped into streams with implicit linear dependencies.
When tasks have dependencies, it is possible to skip polling progress for tasks whose dependent tasks are not yet completed. While implementing a general-purpose task management system to track dependencies is complex, we will demonstrate in \ref{eg:async-class} through an example how users can manage task dependencies within their \pollfn.

\subsubsection{Poll Function Overhead}
In addition to the number of pending tasks, the overhead of individual progress poll functions (\pollfn) can also affect the average event response latency. If a single \pollfn\ takes a disproportionate amount of time to execute, the overall response time to events will increase. This effect is illustrated in Figure~\ref{fig:poll-overhead}, where we manually inserted delays in \pollfn\ when the task is still pending.

The MPIX Async interface is designed for lightweight \pollfn\ functions and is not suitable for tasks that require significant CPU cycles to respond to an event. This is generally true for all collated progress mechanisms: when one part of the progress takes significant overhead, it negatively impacts the performance of other tasks.

To avoid heavy \pollfn\ overhead, it is recommended to enqueue events and postpone the heavy work outside of the progress callbacks. This approach ensures that the \pollfn\ remains lightweight, minimizing its impact on overall performance.

\subsubsection{Thread Contention}
\begin{figure}
    \begin{minipage}[t]{0.45\textwidth}
        \centering
    \includegraphics[width=\columnwidth]{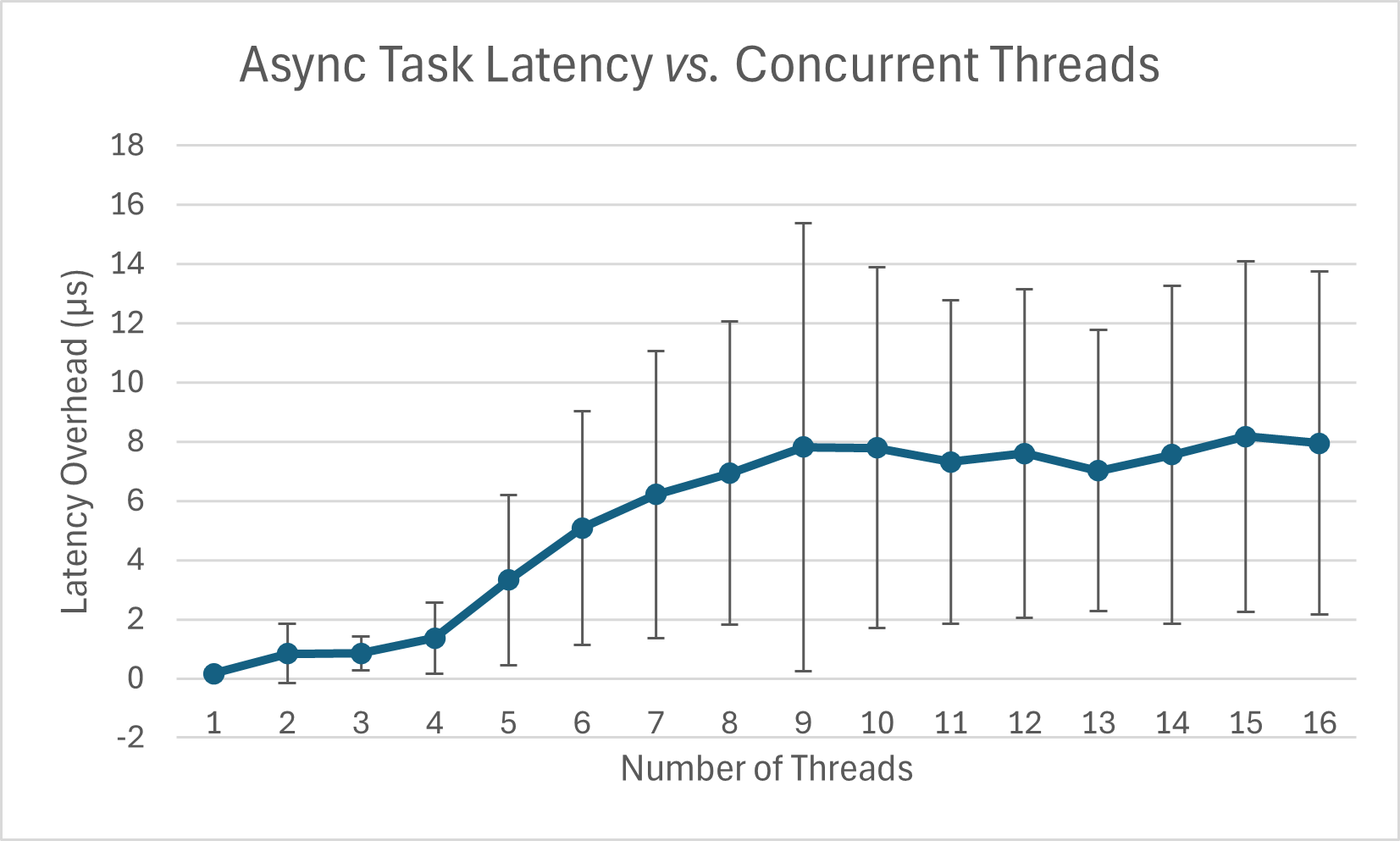}
    \caption{Latency overhead in microseconds as the number of concurrent progress threads increases. Each measurement runs 10 concurrent pending tasks.}
    \label{fig:poll-lock}
    \end{minipage}\qquad
    \begin{minipage}[t]{0.45\textwidth}
        \centering
    \includegraphics[width=\columnwidth]{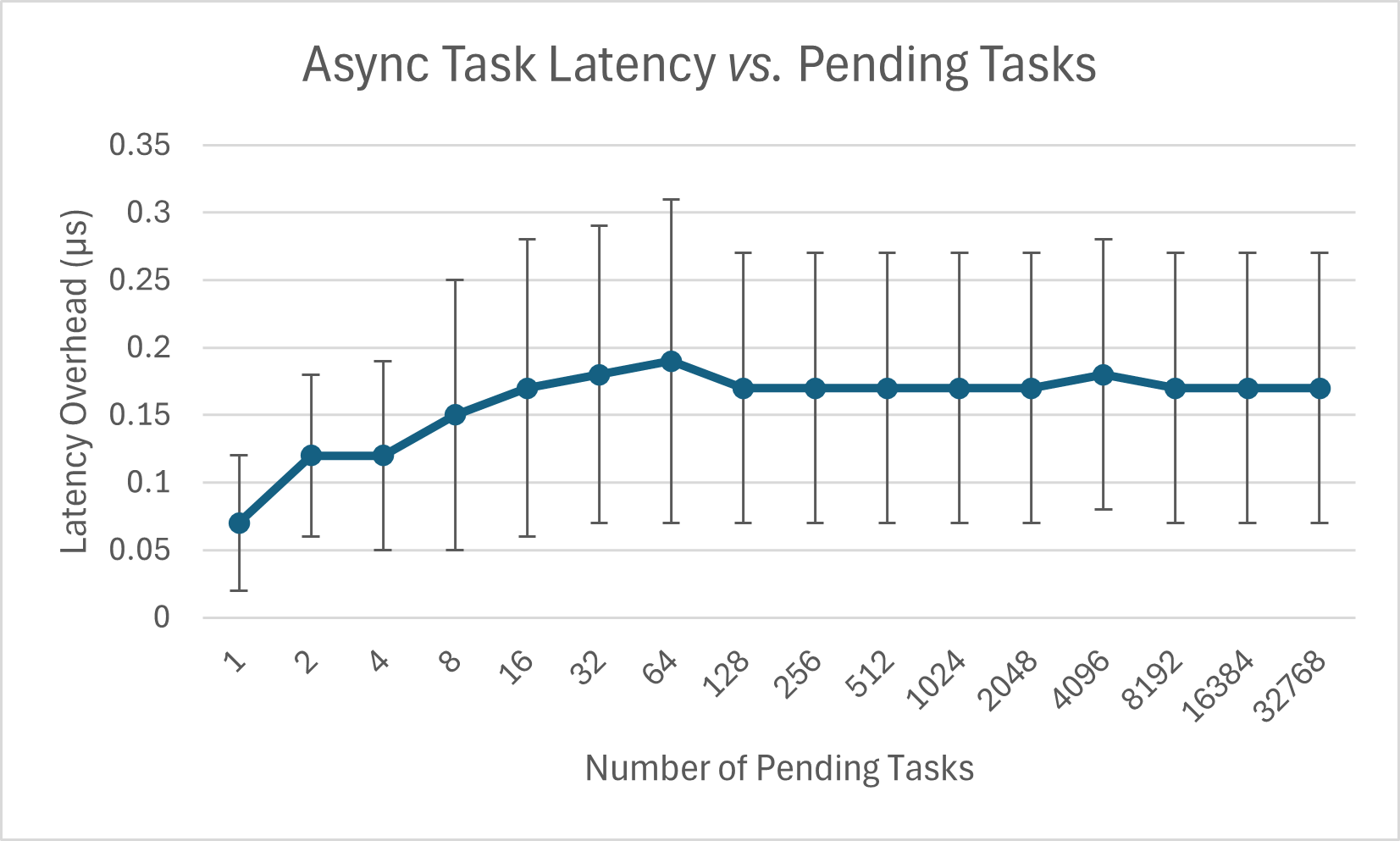}
    \caption{Latency versus the number of pending tasks when the progress callback only checks the task at the top of the queue.}
    \label{fig:task-class}
    \end{minipage}
\end{figure}

When multiple threads concurrently execute progress, they contend for a lock to avoid corrupting the global pending task list.
Even if the tasks are independent, multiple threads running collated progress will still contend for locks, leading to performance degradation. 
As illustrated in Figure~\ref{fig:poll-lock}, the observed latency increases with the number of concurrent progress threads.

It's important to note that individual \pollfn\ functions may access application-specific global states that require lock protection from other parts of the application code. This lock protection should be implemented within the \pollfn\ by the application. MPI only ensures thread safety between MPI progress calls.

To avoid performance degradation, it is advisable to limit the number of progress threads--a single progress thread often suffices. Sometimes, MPI progress loops are invoked implicitly. For example, blocking calls in MPI, such as \texttt{MPI\_Recv}, often include implicit progress similar to \texttt{MPI\_Wait}. Even initiation calls such as \texttt{MPI\_Isend} may contend for a lock with the MPI progress thread. This global lock contention contributes to the notorious poor performance of \texttt{MPI\_THREAD\_MULTIPLE} \cite{mpix-stream}.

Using MPIX Stream appropriately can mitigate the issue of global thread contention. We will demonstrate this in \ref{eg:async-stream}.

\subsection{Async task class} \label{eg:async-class}
To avoid wasting cycles polling progress for tasks with pending dependencies, it is essential to track task dependencies.
This can be managed either within MPI implementations or inside the callback \pollfn.
However, MPI implementations can only handle general task graphs with potentially unbounded complexity, while applications often have much simpler dependency structures.
Therefore, it is simpler and more effective for applications to manage task dependencies within the \pollfn.
Instead of polling progress for individual asynchronous tasks, users can design task subsystems or asynchronous task classes. Within the \pollfn, they can poll progress for the entire task class.

In Listing~\ref{code:task-class}, instead of calling \texttt{MPIX\_Async\_start} for each task individually, the tasks are enqueued and managed within the application.
A single callback, \texttt{class\_poll}, is employed to manage the entire task queue.
For simplicity, we assume all tasks are to be completed in order. 
Therefore, \texttt{class\_poll} only needs to check the completion status of the task at the top of the queue.

\begin{lstlisting}[style=example, caption=An example code for managing an entire task queue, label=code:task-class]
struct task {
    double wtime_end;
    struct task *next;
};

struct task *head, *tail;

/* NOTE: if tasks are to be added and progressed from multiple
   threads, the global task queue will need lock protection.
 */
static int class_poll(MPIX_Async_thing thing)
{
    double tm = MPI_Wtime();
    while (head && tm >= head->wtime_end) {
        struct task *p = head;
        head = head->next;
        free(p);
    }
    if (head == NULL) {
        return MPIX_ASYNC_DONE;
    } else {
        return MPIX_ASYNC_NOPROGRESS;
    }
}

static void add_async(void)
{
    struct task *p = malloc(sizeof(struct task));
    p->wtime_end = MPI_Wtime() + INTERVAL;
    p->next = NULL;

    if (head == NULL) {
        head = p;
        tail = p;
    } else {
        /* simply append */
        tail->next = p;
        tail = p;
    }
}

int main(int argc, const char **argv)
{
    MPI_Init(NULL, NULL);

    int count = 10;
    for (int i = 0; i < count; i++) {
        add_async();
    }
    MPIX_Async_start(class_poll, head);
    while (head) MPIX_Stream_progress(MPIX_STREAM_NULL);

    MPI_Finalize();
    return 0;
}
\end{lstlisting}

As shown in Figure~\ref{fig:task-class}, the average latency stays constant (within measurement noise) regardless of the number of pending tasks.

\subsection{Concurrent progress streams} \label{eg:async-stream}
If running multiple progress threads is necessary, or if there are multiple task systems where collating progress might affect each other's latency, one should consider using MPIX Stream to avoid contention between the various threads.

Listing~\ref{eg:async-stream} demonstrates the use of MPIX streams to scale up the application with multiple threads. We create a separate MPIX stream for each thread. Each thread uses its own stream in \texttt{MPIX\_Async\_start} and \texttt{MPIX\_Stream\_progress}, thereby avoiding lock contention between threads.
\begin{lstlisting}[style=example, caption=An example code using MPIX Streams to manage multiple progress threads , label=code:stream]
#define NUM_TASKS 10

struct dummy_state {
    double wtime_complete;
    int *counter_ptr;
};

static int dummy_poll(MPIX_Async_thing thing)
{
    struct dummy_state *p = MPIX_Async_get_state(thing);
    if (MPI_Wtime() >= p->wtime_complete) {
        (*(p->counter_ptr))--;
        free(p);
        return MPIX_ASYNC_DONE;
    }
    return MPIX_ASYNC_NOPROGRESS;
}

static void add_async(int *counter_ptr, MPIX_Stream stream)
{
    struct dummy_state *p = malloc(sizeof(struct dummy_state));
    
    p->wtime_complete = MPI_Wtime() + INTERVAL + rand() * 1e-5 / RAND_MAX;
    p->counter_ptr = counter_ptr;
    MPIX_Async_start(dummy_poll, p, stream);
}

MPIX_Stream streams[MAX_THREADS];

void *thread_fn(void *arg)
{
    int thread_id = (int) (intptr_t) arg;
    MPIX_Stream stream = streams[thread_id];
    int counter = NUM_TASKS;
    for (int i = 0; i < NUM_TASKS; i++) {
        add_async(&counter, stream);
    }
    while (counter > 0) MPIX_Stream_progress(stream);
}

int main(int argc, const char **argv)
{
    int provided;
    MPI_Init_thread(NULL, NULL, MPI_THREAD_MULTIPLE, &provided);
    assert(provided == MPI_THREAD_MULTIPLE);

    int num_threads = 10;
    for (int i = 0; i < num_threads; i++) {
        MPIX_Stream_create(MPI_INFO_NULL, &streams[i]);
    }

    pthread_t thread_ids[MAX_THREADS];
    for (int i = 0; i < num_threads; i++) {
        pthread_create(&thread_ids[i], NULL,
                       thread_fn, (void *) (intptr_t) i);
    }

    for (int i = 0; i < num_threads; i++) {
        pthread_join(thread_ids[i], NULL);
    }

    for (int i = 0; i < num_threads; i++) {
        MPIX_Stream_free(&streams[i]);
    }
    MPI_Finalize();

    return 0;
}

\end{lstlisting}

\begin{figure}
    \centering
    \begin{minipage}[t]{0.45\textwidth}
        \centering
    \includegraphics[width=\textwidth]{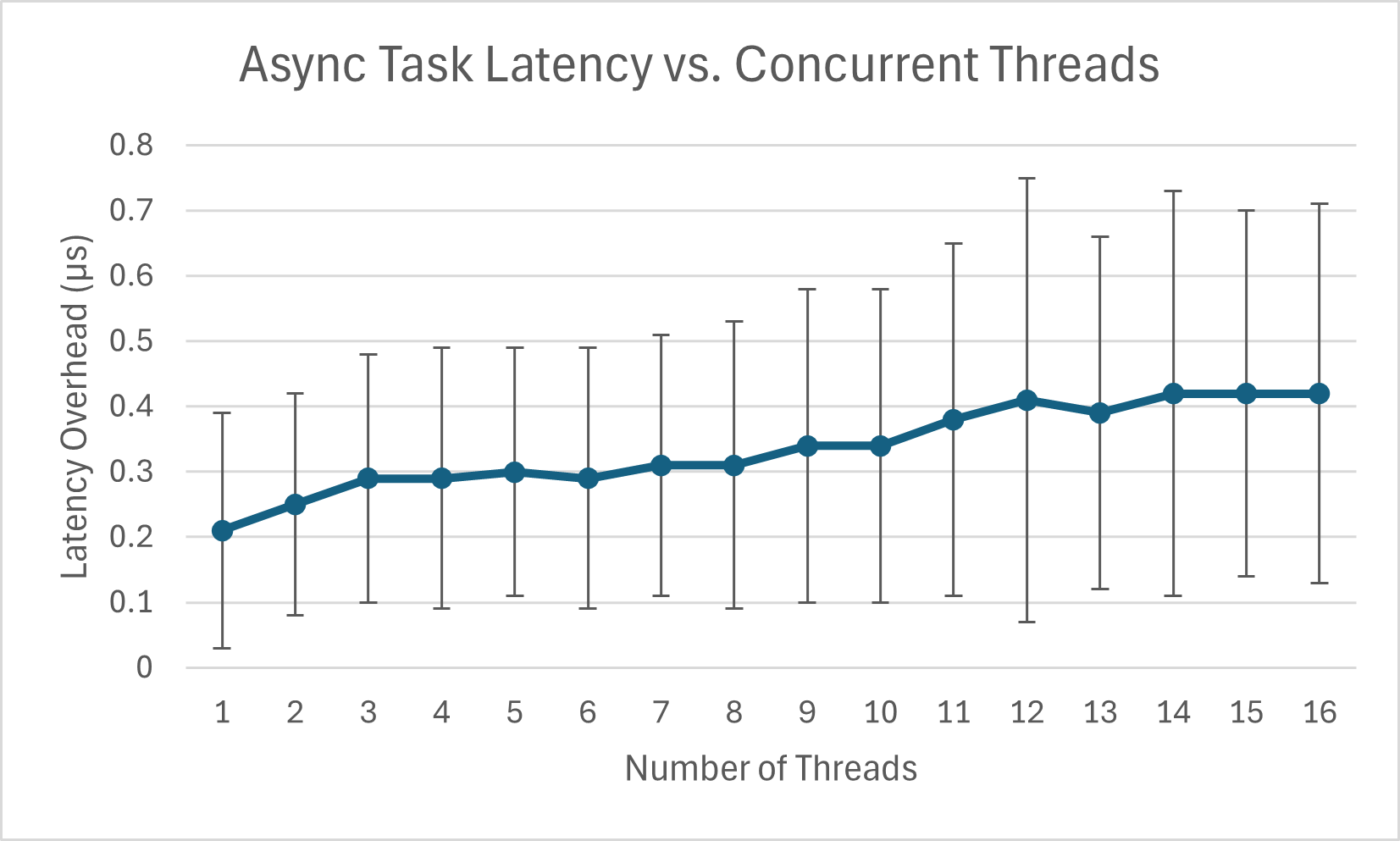}
    \caption{Latency versus the number of concurrent progress threads using different MPIX streams. Each measurement runs 10 concurrent pending tasks.}
    \label{fig:task-stream}
    \end{minipage}\qquad
    \begin{minipage}[t]{0.45\textwidth}
        \centering
        \includegraphics[width=\textwidth]{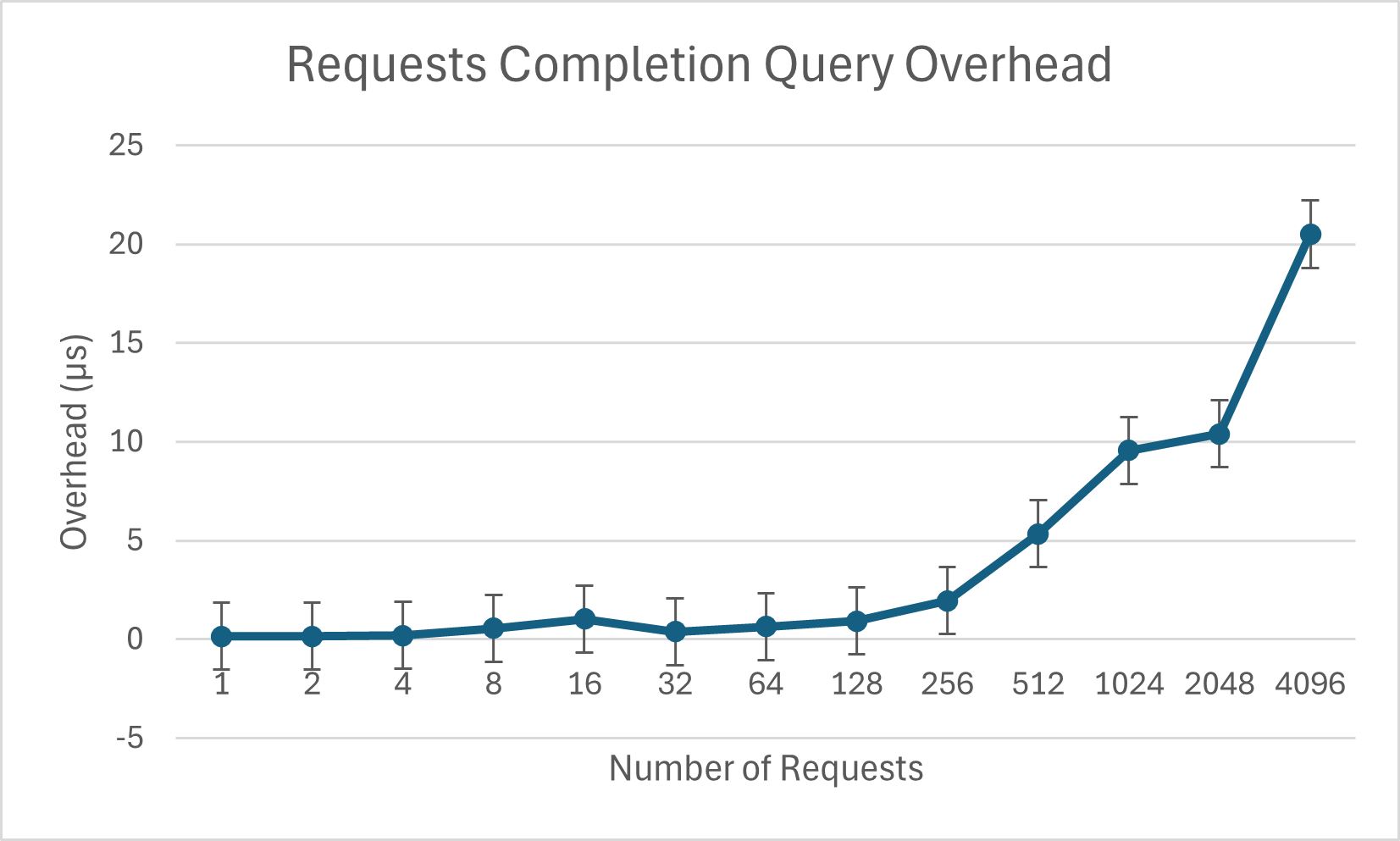}
        \caption{Overhead of generating request completion events via explicit queries.}
        \label{fig:event}
    \end{minipage}\qquad
    
\end{figure}

As shown in Figure~\ref{fig:task-stream}, the average progress latency does not increase significantly as the number of threads increases. The slight increase in latency is within measurement noise and is likely attributable to core power fluctuations due to the number of active cores.

\subsection{Request Completion Callback} \label{eg:async-event}
While the MPIX Async API does not directly provide event callbacks for an event-driven programming style, generating events within the progress hook (i.e., \pollfn) is straightforward.
The following example demonstrates using the MPIX Async API and \texttt{MPIX\_Request\_is\_complete} to generate MPI request completion events.

\begin{lstlisting}[style=example, caption=An example of implementing callbacks upon completion of MPI requests, label=code:event]

int num_requests;
MPI_Request request_array[MAX_REQUESTS];

static int dummy_poll(MPIX_Async_thing thing)
{
    int num_pending = 0;
    for (int i = 0; i < num_requests; i++) {
        if (request_array[i] == MPI_REQUEST_NULL) {
            /* NOOP */
        } else if (MPIX_Request_is_complete(request_array[i])) {
            /* call on-completion actions */
            MPI_Request_free(&request_array[i]);
        } else {
            num_pending++;
        }
    }
    if (num_pending == 0) {
        return MPIX_ASYNC_DONE;
    }
    return MPIX_ASYNC_NOPROGRESS;
}
\end{lstlisting}
The MPI requests are completed within MPI's native progress, as shown in Listing~\ref{lst:mpich-progress}. Thus, using a separate query loop to generate request completion events is less efficient than directly generating events within MPI's native progress when the requests are first completed. However, the overhead of \texttt{MPIX\_Request\_is\_complete} is usually just an atomic read instruction, making it acceptable as long as the number of pending requests is not too large.

Figure~\ref{fig:event} measures the overhead of querying requests using the for-loop in Listing~\ref{code:event}. The overhead remains within the measurement noise when there are fewer than 256 pending requests.

\subsection{Synchronization via Generalized Request} \label{eg:grequest}
The MPIX Async API perfectly complements MPI generalized requests. MPIX Async provides a progression mechanism for asynchronous tasks, while MPI generalized requests offer a tracking handle—an MPI request—that can be used with functions like \texttt{MPI\_Wait}. Together, they enable the extension of MPI functionality at the application layer.

The following example demonstrates how MPIX Async is used in conjunction with MPI generalized requests.

\begin{lstlisting}[style=example, caption=An example of using MPI generaized request and MPIX Async, label=code:grequest]
struct dummy_state {
    double wtime_complete;
    MPI_Request greq;
};

static int dummy_poll(MPIX_Async_thing thing)
{
    struct dummy_state *p = MPIX_Async_get_state(thing);
    if (MPI_Wtime() > p->wtime_complete) {
        MPI_Grequest_complete(p->greq);
        free(p);
        return MPIX_ASYNC_DONE;
    }
    return MPIX_ASYNC_NOPROGRESS;
}

static int query_fn(void *extra_state, MPI_Status *status) { return MPI_SUCCESS; }
static int free_fn(void *extra_state) { return MPI_SUCCESS; }
static int cancel_fn(void *extra_state, int complete) { return MPI_SUCCESS; }

void main(void) {
    MPI_Init(NULL, NULL);
    
    MPI_Request greq;
    MPI_Grequest_start(query_fn, free_fn, cancel_fn, NULL, &greq);
    
    struct dummy_state *p = malloc(sizeof(*p));
    p->wtime_complete = MPI_Wtime() + INTERVAL;
    p->greq = greq;
    MPIX_Async_start(dummy_poll, p, MPIX_STREAM_NULL);

    MPI_Wait(&greq, MPI_STATUS_IGNORE);
    MPI_Finalize();
}
\end{lstlisting}

For simplicity, dummy \texttt{query\_fn}, \texttt{free\_fn}, and \texttt{cancel\_fn} functions are used.
Calling \texttt{MPI\_Wait} on a generalized request effectively replaces the manual wait loop shown in Listing~\ref{code:sync}.

\subsection{User-level collective algorithms}
One of the motivations behind the MPIX Async extension is to enable user-level implementation of collective algorithms with performance comparable to native implementations.
Collective algorithms are essentially a collection of communication patterns built on a core set of operations, including MPI point-to-point operations, buffer copies, and local reductions.

A key advantage of native collective implementations is their tight integration with the MPI progress.
MPIX Async is designed to provide the same level of integration to user-level applications.

Below, we present an example of implementing a user-level allreduce algorithm using the MPIX Async APIs.
This example implements the recursive doubling allreduce algorithm\cite{ruefenacht2017generalisation}.
For simplicity, the datatype is restricted to \texttt{MPI\_INT}, the op to \texttt{MPI\_SUM}, and the number of processes to a power of 2.

\begin{figure}
        \centering
        \includegraphics[width=0.5\textwidth]{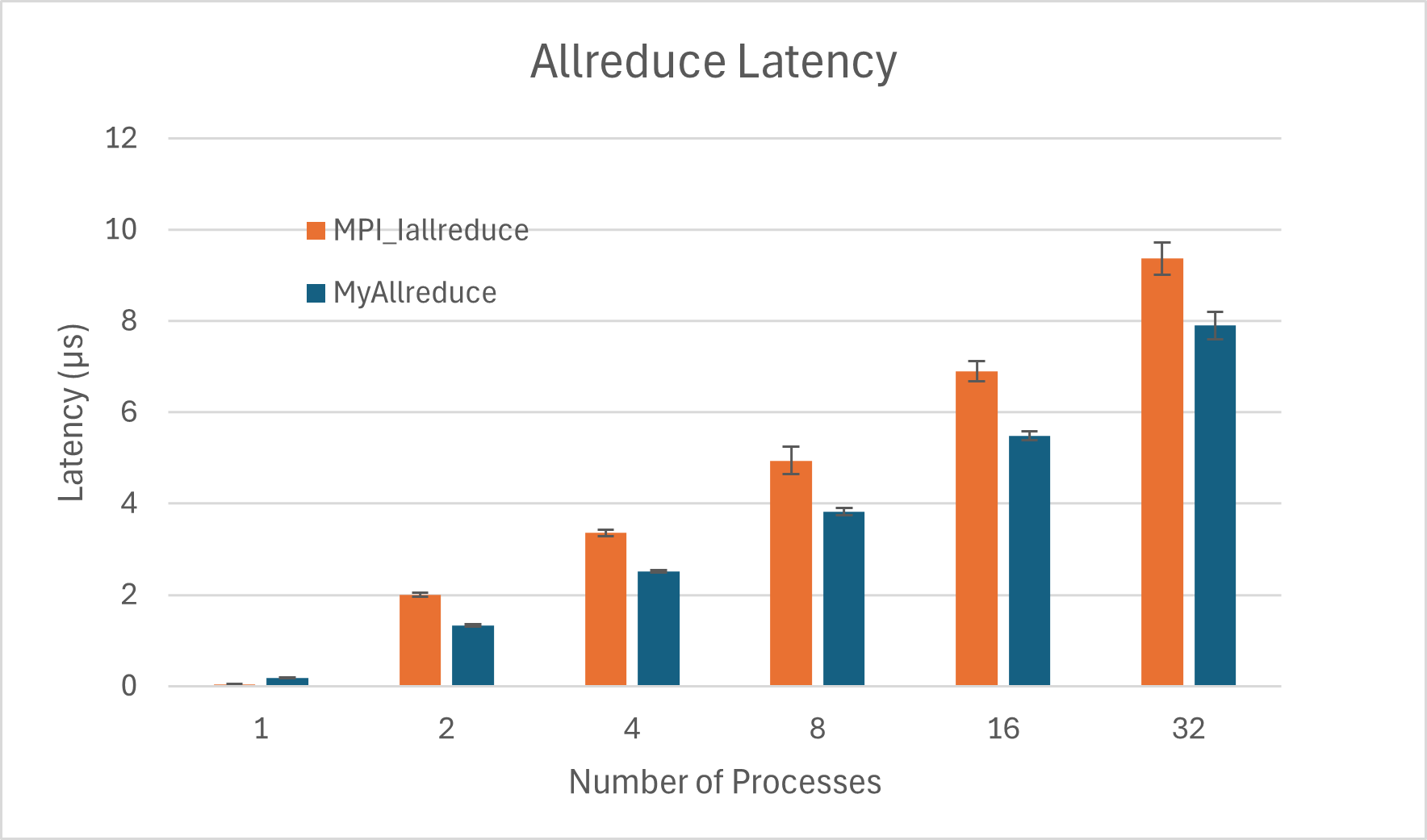}
        \caption{Custom single-integer allreduce latency vs \texttt{MPI\_Iallreduce}. Intel(R) Xeon(R) CPU E5-2695 v4 @ 2.10GHz. Nodes interconnect via Omni-Path Fabric. One process per node.}
        \label{fig:allreduce}
\end{figure}

\begin{lstlisting}[style=example, caption=An example of a user-level allreduce algorithm, label=code:allreduce]
struct myallreduce {
    int *buf, *tmp_buf;
    int count;
    MPI_Comm comm;
    int rank, size;
    int tag;
    int mask;
    MPI_Request reqs[2];  /* send request and recv request for each round */
    bool *done_ptr; /* external completion flag */
};

static int myallreduce_poll(MPIX_Async_thing thing)
{
    struct myallreduce *p = MPIX_Async_get_state(thing);

    int req_done = 0;
    for (int i = 0; i < 2; i++) {
        if (p->reqs[i] == MPI_REQUEST_NULL) {
            req_done++;
        } else if (MPIX_Request_is_complete(p->reqs[i])) {
            MPI_Request_free(&p->reqs[i]);
            req_done++;
        }
    }
    if (req_done != 2) {
        return MPIX_ASYNC_NOPROGRESS;
    }

    if (p->mask > 1) {
        for (int i = 0; i < p->count; i++) {
            p->buf[i] += p->tmp_buf[i];
        }
    }

    if (p->mask == p->size) {
        *(p->done_ptr) = true;
        free(p->tmp_buf);
        free(p);
        return MPIX_ASYNC_DONE;
    }

    int dst = p->rank ^ p->mask;
    MPI_Irecv(p->tmp_buf, p->count, MPI_INT, dst, p->tag, p->comm, &p->reqs[0]);
    MPI_Isend(p->buf, p->count, MPI_INT, dst, p->tag, p->comm, &p->reqs[1]);
    p->mask <<= 1;

    return MPIX_ASYNC_NOPROGRESS;
}

void MyAllreduce(const void *sendbuf, void *recvbuf, int count, MPI_Datatype datatype, MPI_Op op, MPI_Comm comm)
{
    int rank, size;
    MPI_Comm_rank(comm, &rank);
    MPI_Comm_size(comm, &size);

    /* only deal with a special case */
    assert(sendbuf == MPI_IN_PLACE && datatype == MPI_INT && op == MPI_SUM);
    assert(is_pof2(size));

    struct myallreduce *p = malloc(sizeof(*p));
    p->buf = recvbuf;
    p->count = count;
    p->tmp_buf = malloc(count * sizeof(int));
    p->reqs[0] = p->reqs[1] = MPI_REQUEST_NULL;
    p->comm = comm;
    p->rank = rank;
    p->size = size;
    p->mask = 1;
    p->tag = MYALLREDUCE_TAG;

    bool done = false;
    p->done_ptr = &done;

    MPIX_Async_start(myallreduce_poll, p, MPIX_STREAM_NULL);
    while (!done) MPIX_Stream_progress(MPIX_STREAM_NULL);
}
\end{lstlisting}
To compare the performance of this custom user-level allreduce against MPICH's \texttt{MPI\_Iallreduce} using the same recursive doubling algorithm, we conducted experiments on the Bebop cluster at Argonne National Laboratory Computing Resource Center. The experiment measures the latency of allreduce of a single integer. The results are shown in Figure~\ref{fig:allreduce}.

The custom user-level implementation actually outperforms MPICH's native \texttt{MPI\_Iallreduce}. We believe this is due to the specific assumptions and shortcuts in the custom implementation. For example, we assume the number of processes is a power of 2 and that \texttt{sendbuf} is \texttt{MPI\_INPLACE}, which avoids certain checks and branches. Additionally, restricting to \texttt{MPI\_INT} and \texttt{MPI\_SUM} avoids a datatype switch and the function-call overhead of calling an operation function.

This highlights an advantage of custom code over an optimized MPI implementation: the former can leverage specific contexts from the application to avoid complexities and achieve greater efficiency.

\section{Related Work} \label{related}
Several related works address issues with MPI progress and the management of asynchronous tasks within MPI. Here, we provide a brief review and compare these approaches with our proposed methods.

\subsection{MPICH Asynchronous Progress}
With MPICH and its derived MPI implementations, users can set an environment variable, such as \texttt{MPIR\_CVAR\_ASYNC\_PROGRESS}, to enable a dedicated background thread to poll progress in a busy loop. This is the simplest method for ensuring that MPI will always progress and complete any nonblocking operations after they are initiated, without requiring additional MPI calls from the application. However, this method has significant performance issues and is generally not recommended.

The first issue is lock contention between the progress thread and the main thread. Enabling async progress forces the use of the \texttt{MPI\_THREAD\_MULTIPLE} thread level, where all MPI operations require lock protection for thread safety. Because the async progress thread constantly tries to make progress on operations, it creates latency overhead for all MPI calls due to lock contention. On many systems, thread locks are implemented using unfair mutexes, which can lead to a lock monopoly issue. In this scenario, a thread that releases a lock and immediately tries to reacquire it has a better chance of getting the lock than another thread trying to acquire the same lock. Consequently, MPI calls from the main thread may experience disproportionate latency due to the difficulty of acquiring the lock from the async thread. Although adopting a fair mutex implementation may alleviate this issue, the lock contention overhead remains significant, especially for small messages.

Secondly, the async progress option adds one progress thread for every MPI process. In an oversubscribed situation, where there are as many MPI processes on a single node as there are cores, the extra async thread significantly slows down the application due to CPU core sharing among all threads.

MVAPICH \cite{ruhela2019efficient} has proposed a design to address these issues by identifying scenarios where asynchronous progress is required and putting the async thread to sleep when it is not required or beneficial. Their benchmarks showed up to a 60% performance improvement over the original MPICH design.

With the extension \texttt{MPIX\_Stream\_progress}, applications can easily implement the async progress thread within the application layer. The same tuning design as that adopted in MVAPICH can be applied. Instead of the implementation detecting where async progress is required, the application can know where it is needed by design, thus controlling the progress thread more precisely and directly. \texttt{MPIX\_Stream\_progress} also allows applications to use MPIX streams to target async progress for specific contexts, thereby avoiding lock contention issues altogether. In contrast, it is difficult for an MPI implementation to detect user contexts or determine the scope of global progress.

\subsection{Generalized Request}
MPI includes an API called generalized requests, which allows users to wrap an asynchronous task into an MPI request, enabling it to be managed similarly to an MPI nonblocking operation.
A generalized request is created via \texttt{MPI\_Grequest\_start}:
\begin{lstlisting}[language=C, frame=tb]
int MPI_Grequest_start(query_fn, free_fn, cancel_fn, extra_state, request)
\end{lstlisting}

MPI interfaces with the user async tasks via three callbacks: \texttt{query\_fn}, \texttt{free\_fn}, and \texttt{cancel\_fn}. \texttt{query\_fn} is called to fill in an MPI status after the request is completed, \texttt{free\_fn} is called when the request is freed, and \texttt{cancel\_fn} is called when the request is canceled.

While the generalized request behaves like a real MPI request--i.e., it can be passed to \texttt{MPI\_Test} or \texttt{MPI\_Wait}--it does not fulfill the primary purpose of users calling \texttt{MPI\_Test} or \texttt{MPI\_Wait}, which is to poke progress on the nonblocking operation behind the request.
There is no built-in progress mechanism for generalized requests; users are expected to progress the async task behind the generalized request outside of MPI.
This raises questions about the usefulness of generalized requests. Since users don't rely on MPI to progress the task or query its status, a generalized MPI request is merely a redundant handle.
Indeed, a large-scale survey of MPI applications \cite{laguna2019large} found no usages of generalized requests. The few works in the literature that use generalized requests are all for implementing MPI APIs\cite{latham2007extending,hatanaka2017offloaded} on top of a core implementation, necessitating the use of MPI requests. Interestingly, both works highlighted the lack of progress semantics in generalized requests and proposed adding a progress-polling callback to the interface.

We believe the MPIX async extension proposed in this paper perfectly complements the generalized request by providing a progression mechanism. 
We will demonstrate this case with an example in the following section~\ref{eg:grequest}.

\subsection{MPIX Schedule}
MPIX Schedule\cite{schafer2019user} is a proposal to expose MPI's internal nonblocking collective system to applications so that users can create their own nonblocking collective algorithms or a series of coordinated MPI operations similar to a nonblocking MPI collective.
The proposal includes the following APIs:

\begin{lstlisting}[language=C, frame=tb]
int MPIX_Schedule_create(MPIX_Schedule *schedule, int auto_free);

int MPIX_Schedule_add_operation(MPIX_Schedule schedule, MPI_Request request, int auto_free);

int MPIX_Schedule_add_mpi_operation(MPIX_Schedule schedule, MPI_Op op, void *invec, void *inoutvec, int len, MPI_Datatype datatype);

int MPIX_Schedule_mark_reset_point(MPIX_Schedule schedule);

int MPIX_Schedule_mark_completion_point(MPIX_Schedule schedule);

int MPIX_Schedule_create_round(MPIX_Schedule schedule);

int MPIX_Schedule_commit(MPIX_Schedule schedule, MPI_Request *request);

int MPIX_Schedule_free(MPIX_Schedule *schedule);
\end{lstlisting}

These APIs are designed for a specific implementation of persistent nonblocking collectives, which includes a set of operations for initiation, a set of operations for finalization, and rounds of operations for the repeated invocation of the algorithm.

However, it is challenging for MPIX Schedule to accommodate non-MPI operations. The proposal only has APIs to add operations represented as an MPI request or an MPI op. While it is possible to wrap a custom asynchronous operation via a generalized request, the usage is cumbersome. More critically, the lack of a progress mechanism limits the performance users can achieve compared to what is possible within an MPI implementation.

In contrast, the MPIX Async APIs address the root issue of providing interoperable MPI progress. Its simple yet flexible progress hook-based interface can accommodate arbitrary asynchronous tasks. Combined with generalized requests, it enables users to effectively experiment with MPI extensions, such as experimental collective algorithms, entirely from the application layer.

\subsection{MPIX Continue}
MPIX Continue \cite{schuchart2021callback} is a proposal that allows MPI to call back a user-defined function upon the completion of a request, eliminating the need for the application to continuously test the request for completion.

The proposal consists of the following APIs:
\begin{lstlisting}[language=C, frame=tb]
int MPIX_Continue_init(MPI_Request *cont_req, MPI_Info info);

int MPIX_Continue(MPI_Request *op_request, int *flag, MPIX_Continue_cb_function *cb, void *cb_data, MPI_Status *status, MPI_Request cont_req);

int MPIX_Continueall(int count, MPI_Request op_request[], int *flag, MPIX_Continue_cb_function *cb, void *cb_data, MPI_Status statuses[], MPI_Request cont_req);
\end{lstlisting}

The motivation behind the MPIX Continue APIs is to simplify the management of MPI requests within a task-based programming model. Task-based programming models often have their own task tracking, scheduling, and progression mechanisms. However, MPI's nonblocking interface requires applications to actively test or wait on individual requests, which can lead to thread contention and wasted CPU cycles due to redundant progression.

Our proposed extensions address directly many of the challenges that the MPIX Continue proposal seeks to resolve.
\texttt{MPIX\_Stream\_progress} can be used to implement a polling service that integrates with the task runtime system without requiring synchronization of MPI requests.
Meanwhile, \texttt{MPIX\_Request\_is\_complete} allows tasks to query the status of their dependent requests without triggering redundant progress invocations.

The MPIX Continue proposal adopts an event-driven callback interface.
As shown in section~\ref{eg:async-event}, MPIX Async extension can be used to implement a poor man's version of callbacks on request completion.
It is less efficient than the MPIX Continue proposal since the latter can be implemented within an implementation's internal progress rather than active queries in a separate progress hook. However, the overhead should be negligible until the number of registered MPI requests becomes significant.

On the other hand, because the MPIX Continue is integrated within the MPI progress engine, there can be issues, including callback latency interference, thread contention, and potential recursive context, that are complex and difficult to address due to the opaqueness of MPI progress. In comparison, MPIX Async extension provides the explicitness and flexibility for applications to address such issues.
\section{Summary} \label{summary}
In summary, we present a suite of MPI extensions crafted to provide an interoperable MPI progress, which grants applications explicit control over MPI progress management, the ability to incorporate user-defined asynchronous tasks into MPI, and seamless integration with task-based and event-driven programming models. The \texttt{MPIX\_Stream\_progress} and \texttt{MPIX\_Request\_is\_complete} APIs effectively decouple the context for invoking MPI progress and querying the completion status of MPI requests, thereby circumventing synchronization complexities and task-engine interference. Meanwhile, the MPIX Async proposal empowers applications to integrate custom progress hooks into MPI progress, enabling them to harness MPI progress and extend MPI functionality from the user layer. 
Through examples and micro-benchmark testing, we demonstrate the effectiveness of these extensions in bringing MPI to modern asynchronous programming.

% \input{discussion}
% \input{code/allreduce}

% \begin{credits}
\subsubsection{\ackname}
This research was supported by the U.S. Department of Energy, Office of Science, under Contract DE-AC02-06CH11357. 
%We acknowledge funding from the U.S. Department of Energy’s Office of Advanced Scientific Computing Research (ASCR) through S4PST: Stewardship for Programming Systems and Tools, and PESO: Partnering for Scientific Software Ecosystem Stewardship Opportunities.
%We acknowledge support from the Argonne Leadership Computing Facility (ALCF) at Argonne National Laboratory.

% \subsubsection{\discintname}
% The authors have no competing interests to declare that are relevant to the content of this article.
% \end{credits}

\bibliographystyle{splncs04}
\bibliography{references}

\end{document}